\documentclass[fleqn,usenatbib]{mnras}

\usepackage{newtxtext,newtxmath}
\usepackage[T1]{fontenc}

\DeclareRobustCommand{\VAN}[3]{#2}
\let\VANthebibliography\thebibliography
\def\thebibliography{\DeclareRobustCommand{\VAN}[3]{##3}\VANthebibliography}

\usepackage{graphicx}	% Including figure files
\usepackage{CJK}        % To provide Chinese characters
\usepackage[utf8]{inputenc} % For handling Chinese characters
\usepackage{xspace} % Adds \xspace macro that intelligently adds a space, for custom commands.
\usepackage{siunitx} % To handle unit formatting. (\SI{}{})
\usepackage{bpchem} % To handle entering chemical formulae. (\BPChem{})
\usepackage{varioref} % To handle formatting references

\labelformat{chapter}{Chapter~#1}
\labelformat{section}{Section~#1}
\labelformat{appendix}{Appendix~#1}
\labelformat{subsection}{Section~#1}
\labelformat{subsubsection}{Section~#1}
\labelformat{figure}{Figure~#1}
\labelformat{subfigure}{Figure~\thefigure #1}
\labelformat{table}{Table~#1}
\labelformat{equation}{Equation~(#1)}

\hypersetup{breaklinks=true}

\newcommand{\varalpha}{\TextOrMath{\(\Delta\alpha/\alpha\)}{\Delta\alpha/\alpha}\xspace}
\newcommand{\sigsys}{\TextOrMath{$\sigma_{**}$}{\sigma_{**}}\xspace}
\newcommand{\Teff}{\TextOrMath{T$_\textrm{eff}$}{\textrm{T}_\textrm{eff}}\xspace}
\newcommand{\Teffnom}{\TextOrMath{$\mathcal{T}^{\mathrm{N}}_{\mathrm{eff}\odot}$}{\mathcal{T}^{\mathrm{N}}_{\mathrm{eff}\odot}}}
\defcitealias{Berke2022a}{B22a}
\DeclareSIUnit\angstrom{\text{\AA}}
\DeclareSIUnit\bar{bar}

\title[Probing $\alpha$ with solar twins: systematic errors]{Probing Galactic variations in the fine-structure constant using solar twin stars: systematic errors}

\author[D. Berke et al.]{
Daniel A. Berke,$^1$\thanks{\textit{Gemini Observatory North, 670 N. A$\!$`oh\=ok\=u Place, Hilo, HI 96720, USA} Email: daniel.berke@noirlab.edu}
Michael T. Murphy,$^1$
Chris Flynn,$^1$
Fan Liu (刘凡)$^1$
\\
$^{1}$Centre for Astrophysics and Supercomputing, Swinburne University of Technology, Hawthorn, VIC 3122, Australia
}

\date{Accepted XXX. Received YYY; in original form ZZZ}

\pubyear{2022}

\begin{document}
\begin{CJK*}{UTF8}{gbsn}
\label{firstpage}
\pagerange{\pageref{firstpage}--\pageref{lastpage}}
\maketitle

\begin{abstract}
Sun-like stars are a new probe of variations in the fine-structure constant, $\alpha$, via the solar twins approach: velocity separations of close pairs of absorption lines are compared between stars with very similar stellar parameters, i.e.\ effective temperature, metallicity and surface gravity within \SI{100}{\kelvin}, \SI{0.1}{dex} and \SI{0.2}{dex} of the Sun's values.
Here we assess possible systematic errors in this approach by analysing $\gtrsim$10,000 archival exposures from the High-Accuracy Radial velocity Planetary Searcher (HARPS) of 130 stars covering a much broader range of stellar parameters.
We find that each transition pair's separation shows broad, low-order variations with stellar parameters which can be accurately modelled, leaving only a small residual, intrinsic star-to-star scatter of 0--\SI{33}{\meter\per\second} (average $\approx$\SI{7}{\meter\per\second}, $\approx$\SI{1e-4}{\angstrom} at \SI{5000}{\angstrom}).
This limits the precision available from a single pair in one star.
We consider potential systematic errors from a range of instrumental and astrophysical sources (e.g.\ wavelength calibration, charge transfer inefficiency, stellar magnetic activity, line blending) and conclude that variations in elemental abundances, isotope ratios and stellar rotational velocities may explain this star-to-star scatter.
Finally, we find that the solar twins approach can be extended to solar analogues -- within \SI{300}{\kelvin}, \SI{0.3}{dex}, and \SI{0.4}{dex} of the Sun's parameters -- without significant additional systematic errors, allowing a much larger number of stars to be used as probes of variation in $\alpha$, including at much larger distances.
\end{abstract}


%%%%%%%%%%%%%%%%%%%%%%%%%%%%%%%%%%%%%%%%%%%%%%%%%%

%%%%%%%%%%%%%%%%% BODY OF PAPER %%%%%%%%%%%%%%%%%%

\section{Introduction} \label{secondpaper:section:introduction}
Modern theories of physics rely on a set of quantities known as the fundamental constants.
However, we currently have no deeper theory from which to calculate the values of these constants; they can only be measured in nature.
Empirical measurements have so far been consistent with no variation in these quantities, but there is no \textit{a priori} reason for assuming that they are constant.
Without a theory explaining their values, it is essential to test experimentally for variations in them.

We focus here on the fine-structure constant, traditionally denoted $\alpha$ and defined as $\alpha\equiv e^2/\hbar c$.
This paper serves as a companion paper to \citet[][hereafter \citetalias{Berke2022a}]{Berke2022a}, wherein we describe a novel ``solar twins method'' to empirically constrain variation in $\alpha$.
\citetalias{Berke2022a} covers the general theory of the method, while this paper details all the potential sources of systematic error identified and our efforts to avoid or suppress them.

The fine-structure constant characterizes the strength of electromagnetic interactions, making it easy to search for variation in its value directly through spectroscopy.
Experimental searches for variation in $\alpha$ extend back at least to \citet{Savedoff1956}, and have been performed using a variety of methods in the decades since \citep[see e.g.][for recent reviews]{Uzan2011, Martins2017}.
While searches have yet to yield an unambiguous detection of variation, multiple hypothetical extensions to the Standard Model have been and continue to be proposed which include variation in $\alpha$ as a function of some parameter other than the characteristic energy-scale \citep[e.g.,][]{Brans1961, Forgacs1979, Bekenstein1982, Olive2002, Davoudiasl2019}.

In this paper we focus on astronomical searches for variation in $\alpha$ using spectroscopy,
\begin{equation}
\Delta\alpha/\alpha\equiv\frac{\alpha_\text{obs}-\alpha_0}{\alpha_0},
\end{equation}
with $\alpha_0$ being the value measured in the laboratory and $\alpha_\text{obs}$ the value in the object being used a probe.
Spectroscopic searches rely on the fact that the value of $\alpha$ influences the energy level of each atomic orbital state to a different degree.
A change in $\alpha$ would therefore be observable as a change in the wavelengths of transitions between orbital states.

The simplest method involves comparing a transition's measured energy with its value in the laboratory, and has been in use in astronomical spectroscopy since the work of \citet{Savedoff1956}, who calculated a constraint of $\varalpha=(1.8\pm1.6)\times10^{-3}$ in the radio source Cygnus A using the fine-structure splitting doublet of two atomic species.
There are, however, significant systematic errors in this approach which limit its precision.
Errors in measuring the wavelengths of transitions in the laboratory and astrophysical probe all compound to reduce the precision of the constraints on \varalpha that can be obtained.

An improvement in the precision reachable came in 1999 with the development of the many-multiplet (MM) method \citep{Webb1999, Dzuba1999b}, which uses constraints calculated from multiple species and transitions in the same source.
Much work has been done using the MM method with quasar absorption systems in the past two decades \citep[see e.g.,][among others]{Webb1999,Webb2001,Murphy2003,Murphy2004,Kotus2017}, with the current most precise astronomical constraints on \varalpha coming from its use in \citet{Murphy2022}.
They calculated a constraint of $\langle\varalpha\rangle_{\text{W}}=-0.5\pm0.4_\mathrm{stat}\pm0.5_\mathrm{sys}\times10^{-6}$ at redshifts of 0.5--2.4, an improvement of three orders of magnitude compared to the pioneering results of \citet{Savedoff1956}.
However, the level of precision which can be reached with quasar absorption systems on current \SI{10}{\meter}-class telescopes without prohibitively long observing times has been essentially exhausted \citep{Kotus2017}.
Next-generation \SI{30}{\meter}-class telescopes will potentially be able to increase the precision level associated with quasar absorption systems, but in the interim it makes sense to consider alternative spectral sources.

Stars have long offered a tempting probe for searches of variation in $\alpha$.
In contrast to quasar absorption systems which typically contain $\sim10$ usable absorption features, certain stellar spectral types have hundreds or thousands of potentially usable features.
In addition, stars brighter than the brightest quasar are abundant.
However, the complexity of stellar atmospheric physics has generally discouraged the use of stars as probes of \varalpha to date.
Relatively few studies have used stars, specifically white dwarfs \citep{Berengut2013,Hu2020} and red giants \citep{Hees2020}.
However, none of these studies surpassed the precision from quasar absorption systems in \citet{Murphy2022} due to significant systematic errors.
For instance, the asymmetries known to be present in the features of stars with convective photospheres \citep{Dravins1982, Dravins2008} introduce systematic errors in the wavelengths of those features at the level of hundreds of \si{\meter\per\second}, an order of magnitude larger than a part-per-million shift in $\alpha$.
In \citetalias{Berke2022a} we describe our new solar twins method for constraining \varalpha using main-sequence stars as probes for the first time; in this paper we quantify and discuss in detail all the potential systematic errors present in the method.
We demonstrate that main sequence Sun-like stars can be viable and valuable probes for constraining \varalpha when taking the appropriate steps to mitigate systematic errors that we have identified.

Both here and in \citetalias{Berke2022a} we use the terms \textsl{solar twin, solar analogue, solar-type star}, and \textsl{Sun-like star}.
While there are no standard definitions for these terms in the literature, we follow general usage \citep[see e.g.][]{CayreldeStrobel1996} in using the terms solar twin, solar analogue, and solar-type star to refer to stars of decreasing similarity to the Sun.
We also use the catch-all term \textsl{Sun-like star} to refer to all of the aforementioned categories where the exact degree of similarity is irrelevant.
For solar twin, solar analogue, and solar-type star, we adopt the following definitions:
\begin{align} \label{equation:star_definitions}
    \text{Solar twin} &= \left\{
    \begin{array}{r@{\,}l}
         \Teffnom&\pm\,\SI{100}{\kelvin},\\
         \mathrm{[Fe/H]_\odot}&\pm\,\SI{0.1}{dex},\\
         \log{g_\odot}&\pm\,\SI{0.2}{dex},\nonumber\\
    \end{array} \right. \\
    \text{Solar analogue} &= \left\{
    \begin{array}{r@{\,}l}
        \Teffnom&\pm\,\SI{300}{\kelvin},\\
        \mathrm{[Fe/H]_\odot}&\pm\,\SI{0.3}{dex},\\
        \log{g_\odot}&\pm\,\SI{0.4}{dex},\\
    \end{array} \right. \\
    \text{Solar-type star} &= \left\{
    \begin{array}{r@{\,}l}
         (b-y)_\odot&\pm\,\SI{0.128}{mag},\\
         \mathrm{[Fe/H]_\odot}&{}^{+\SI{0.45}{dex}}_{\SI{-0.75}{dex}},\\
         \mathrm{M}_{\mathrm{V}\odot}&\pm\,\SI{0.8}{mag},\nonumber\\ 
    \end{array} \right.
\end{align}
where $\Teffnom=\SI{5772}{\kelvin}$ is the nominal solar effective temperature as defined by the International Astronomical Union \citep{Prsa2016}, [Fe/H] is 0 by definition, $\log{g_\odot}=\SI{4.44}{\centi\meter\per\second\squared}$ is the solar surface gravity\footnote{Calculated as $g_{\odot}=\mathcal{(GM)}^\mathrm{N}_\odot/\mathcal{R}^2_\odot$ where $\mathcal{(GM)}^\mathrm{N}_\odot$ is the nominal solar mass parameter and $\mathcal{R}^2_\odot$ the nominal solar radius as defined by the International Astronomical Union \citep{Prsa2016}.}, the solar Str\"omgren color is $(b-y)_\odot=0.403\pm0.013$ \citep{Holmberg2006}, and the solar absolute magnitude is $\mathrm{M}_{\mathrm{V}\odot}=4.83$ \citep{Allen1976}.
The use of different, photometric parameters for the definition of solar-type stars lies in our use of them for the initial selection; put another way, all stars considered in this paper are solar-type stars, from which we then selected the more restrictive categories of solar analogue and solar twin.
The initial asymmetry in metallicity around zero was intended to mirror the skew in metallicity seen in nearby Galactic stars in order to find as many Sun-like stars as possible, but further constraints (detailed in \ref{section:sample_selection}) meant that we ultimately were left with no stars with [Fe/H] $<-0.45$.

This paper is laid out as follows.
\ref{section:star_and_pair_selection} contains details of the stellar sample analysed (a superset of the stars studied in \citetalias{Berke2022a}) and the process of selecting transitions.
\ref{section:analysis} provides details on HARPS (the instrument used for this work) and systematic errors in the data reduction process.
\ref{section:systematic_errors} contains a detailed analysis of all the sources of systematic error present in the solar twins method, and we summarise our conclusions in \ref{section:conclusions}.

\section{Selection of stars and transition pairs} \label{section:star_and_pair_selection}
The solar twins method is described in \citetalias{Berke2022a}, but we provide a brief summary here to provide context to the selection process for stars and transition pairs.
Current methods of constraining variation in $\alpha$ rely on measuring the change in a transition's wavelength (relative to its laboratory value) which a variation in $\alpha$ would cause.
These methods are prone to systematic errors in measuring wavelengths, both in the laboratory and in the probe (with stars' complex atmospheres especially complicating their use).

The solar twins method, in contrast, involves comparing the separations between pairs of transitions across multiple stars which are `similar enough' to account for any remaining systematic differences between them.
Stellar atmospheric parameters such as \Teff, [Fe/H], and $\log{g}$ vary considerably across all stars, however, so in this paper we quantify just how similar stars must be to be usable.
This differential comparison between pairs of transitions across stars (rather than single transitions to laboratory wavelengths) is one of the key features of the solar twins method, as it removes an entire class of systematic errors related to measuring the absolute wavelengths of transitions.
\citetalias{Berke2022a} focuses on a sample of 18 solar twins, stars that are most similar to the Sun and each other, in order to provide the most homogeneous sample to use for reporting the first results of the solar twins method.
In this paper we instead investigate a larger sample of solar-type stars to determine the degree of similarity necessary within which the method works.

\subsection{Stellar sample selection} \label{section:sample_selection}

\begin{figure}
  \begin{center}
    \includegraphics[width=\linewidth]{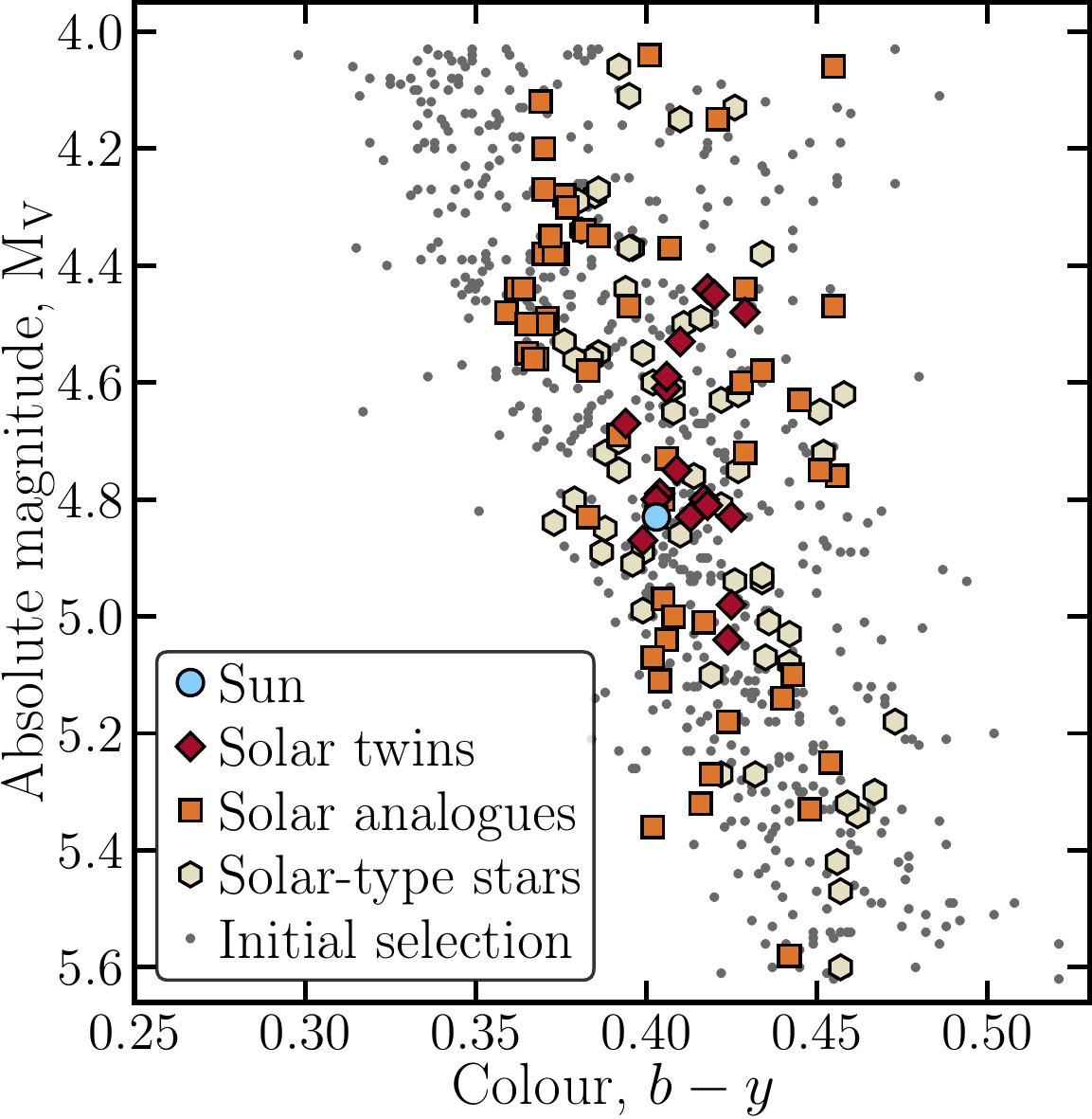}
    \caption{Hertzsprung--Russell diagram of the stars in the sample, comprising 17 solar twins (plus the Sun), 54 solar analogues, and 58 solar-type stars.
    Stars listed as ``initial selection'' were not part of the final sample for reasons explained in the text, but provide context showing the main sequence.
    Colour and absolute magnitude data were taken from \citet{Nordstrom2004}.
    }
    \label{figure:HR-diagram}
  \end{center}
\end{figure}

To acquire a sample of Sun-like stars whose atmospheric parameters are well known (so that we can model their effects on transition pair separation), we used the Geneva--Copenhagen Survey (GCS).
The GCS is an all-sky, magnitude-limited and kinematically-unbiased sample of 14139 nearby F and G dwarfs \citep{Nordstrom2004}, with updated values for their stellar parameters provided by \citet{Casagrande2011}.
The GCS contains information on metallicity, absolute magnitude derived from \textsl{Hipparcos} data in the Johnson-Cousins V band \citep{Johnson1953}, colour in the Str\"omgren colour system \citep{Stroemgren1963}, and effective temperature for each star, with \citet{Casagrande2011} adding surface gravity.
To select ``solar-type stars'' defined in \ref{equation:star_definitions}, we followed the approach of \cite{Datson2012} by making an initial photometric selection using the Str\"omgren $b-y$ colour, absolute magnitude $\mathrm{M}_{\mathrm{V}\odot}$, and metallicity, using the definition of ``solar-type star'' in \ref{equation:star_definitions}.
This initial selection, containing 711 stars, is shown in a Hertzsprung--Russell diagram in \ref{figure:HR-diagram}.

For this work we required high-resolution, precisely-calibrated, high-SNR spectra, for which we used archival spectra from the High Accuracy Radial-velocity Planet Searcher (HARPS) spectrograph (described in more detail in \ref{section:HARPS}).
In order to work with a photon-noise limited sample, we selected only stars which had observations with $200\leq\mathrm{SNR}\leq400$ per \SI{0.8}{\kilo\meter\per\second} pixel in the HARPS archive\footnote{\url{http://archive.eso.org/wdb/wdb/adp/phase3_spectral/form}}.
Visual inspection of spectra with very high $\mathrm{SNR}$ ($\ga450$) showed the presence of artefacts (such as repeated sharp-sided, flat-topped regions of highly elevated flux) which we interpreted as evidence of CCD saturation, so we implemented the upper limit as a precaution.
Of the initial selection of 711 stars, 201 had at least a single observation in the HARPS archive fulfilling these criteria.

Our analysis suggested additional selection criteria which further reduced the number of stars we ultimately used.
Investigation of HARPS wavelength calibration showed that using recently updated calibration files (explained in \ref{section:pixel_column_width}) was crucial to correct for systematic errors at the level of \SI{50}{\meter\per\second} across the entire spectrum \citep{Coffinet2019}.
The relevant correction was not implemented for spectra in the HARPS archive at the time of retrieval, and updated calibration files were only available for a subset of observations dates at the time of our analysis (C. Lovis, priv. comm.).
This reduced the sample size to 164 stars.
In the course  of our analysis we found that stars either significantly hotter than, or metal-poor, compared to the Sun showed increased scatter in pair separations (see \ref{section:stellar_parameter_dependence} for details).
We therefore removed an additional 34 stars with $\Teff>\SI{6072}{\kelvin}$ or $\mathrm{[Fe/H]}<-0.45$.
Additionally, for reasons explained in \ref{section:telluric_absorption_features}, we required stars to have a barycentric radial velocity of within \SI{\pm70}{\kilo\meter\per\second} to facilitate avoiding blends with telluric absorption features, though this ultimately only affected 6 stars which had already been removed.
The final sample comprised 130 stars, for which we were able to obtain 10126 HARPS archival spectra observed between 2004 and 2017.
The number of observations per star ranged from 1 to 3360, with a median value of 9 (14 stars had more than a hundred observations, and 2 had more than a thousand).

\subsection{Selecting transitions} \label{section:transition_selection}
\begin{figure}
  \begin{center}
    \includegraphics[width=80mm]{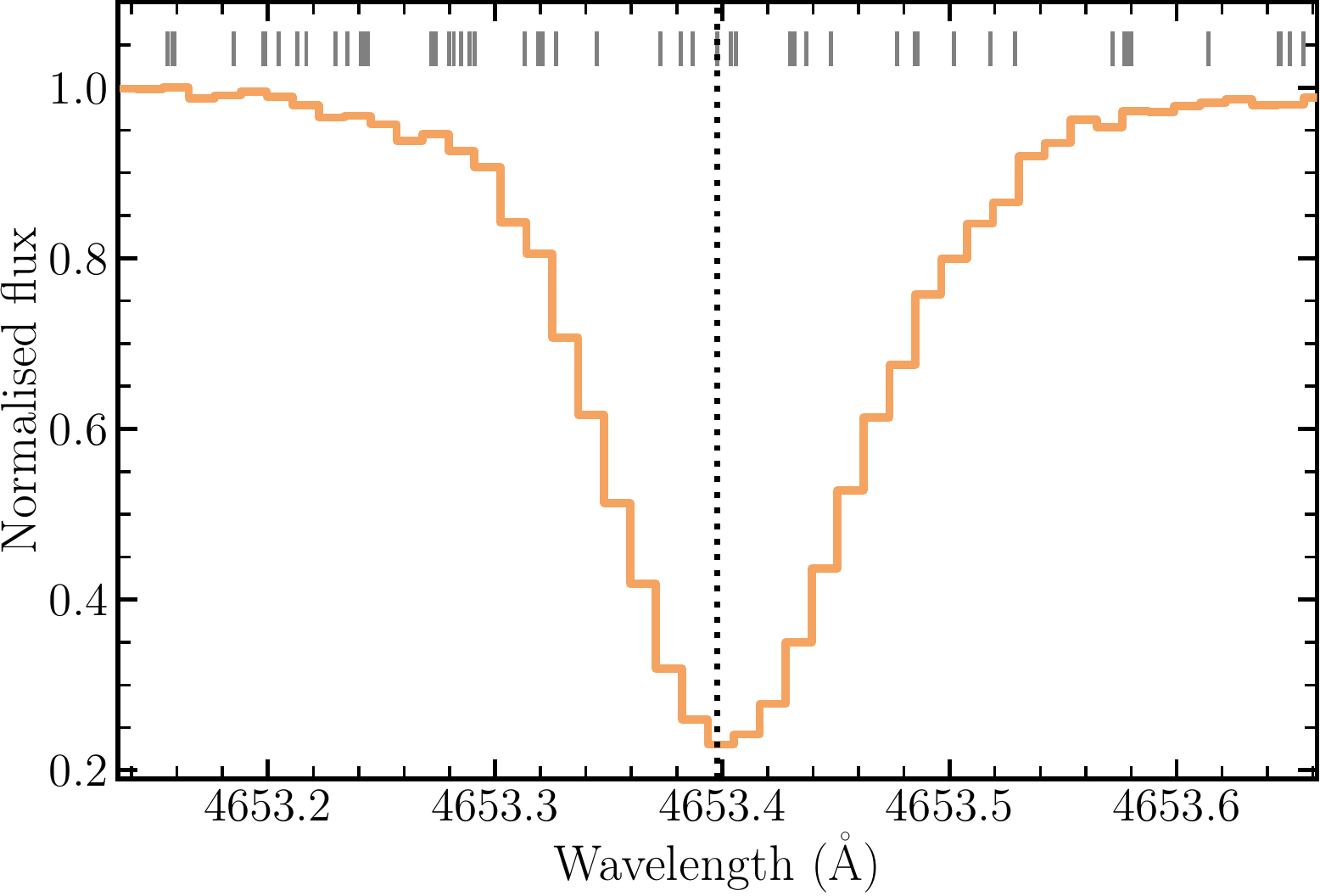}
    \caption{
    Profile of absorption feature in the Sun due to the transition from \ion{Cr}{i} at \SI{4653.460}{\angstrom}.
    The spectrum is from one of the HARPS observations used in this work (taken on 2011-09-29), reflected off Vesta.
    The wavelength scale is in vacuum in the Earth rest frame.
    Wavelengths of 62 transitions from the Kurucz list which fall within this \SI{34}{\kilo\meter\per\second} region of the spectrum (\SI{\sim0.5}{\angstrom}) are plotted as ticks near the top.
    Also visible is the asymmetry of the feature, which may be due to the convective nature of the Sun's photosphere or weak blends with other transitions.
    The dotted vertical line indicates where the fitting procedure we used (described in \citetalias{Berke2022a}, Section 4.1) determined the wavelength of the feature.
    }
    \label{figure:transition_density}
  \end{center}
\end{figure}

For the solar twins method to work, it is critical to match absorption features to the transitions responsible for them, including full information about their upper and lower energy states.
This information is required to calculate the expected change in transition wavelengths for a change in $\alpha$, which is specific to each transition.
\ref{figure:transition_density} shows an example of an absorption feature from one of the 164 transitions in the transition sample defined later in this section.
Note the large number of other transitions (indicated by ticks at the top) that have wavelengths falling close to the one responsible for the majority of the absorption.
This illustrates the importance of verifying exactly which transition is primarily responsible for a given feature.

As a starting point for identifying transitions we used a list of 8843 transitions (A. Lobel, priv. comm.) from the Belgian Repository of fundamental Atomic data and Stellar Spectra (BRASS\footnote{\url{http://brass.sdf.org/}}, \citealt{Laverick2017, Laverick2018b, Laverick2018a}).
These transitions were used by the BRASS project to compute a synthetic solar spectrum and could thus be matched to specific absorption features.
The BRASS list did not contain complete orbital configuration information for transitions, so we cross-matched the BRASS list with the Atomic Spectra Database\footnote{\url{https://www.nist.gov/pml/atomic-spectra-database}} \citep{Kramida1999} of the National Institute of Standards and Technology (NIST) while also requiring that transitions be found in a separate list of transitions from R. Kurucz\footnote{\url{http://kurucz.harvard.edu/linelists/gfnew/}, with the filename ``gfallvac08oct17.dat''.}.
For some transitions the NIST database did not have orbital configurations listed; for those transitions we made use of the following publications for the corresponding ionic species to make the final identification: \ion{Fe}{I}, \citet{Nave1994}; \ion{Co}{I}, \citet{Sugar1981}; \ion{Si}{I}, \citet{Martin1983}; \ion{Ti}{I}, \citet{Saloman2012}.
In order to avoid any features which would be blended at HARPS's resolution of $\mathrm{R}\approx115000$ \citep{Mayor2003}, we removed any transitions within \SI{9.1}{\kilo\meter\per\second} ($3.5\times$ HARPS's resolution element) of other transitions in the BRASS list.

To avoid errors with measuring the centroids of weak or saturated features, we selected only transitions whose features had normalized depths of between 0.15 and 0.90 relative to the local continuum in the Sun.
We measured the normalized depth of each feature in the BRASS list in a high-SNR HARPS observation (\(\text{SNR}=316.8\)) of the solar spectrum reflected off Vesta.
This was done by fitting an absorption feature at the wavelength of each transition using a four-parameter integrated Gaussian function (for details of the automated fitting process, see \citetalias{Berke2022a}).

\subsubsection{Blending with telluric absorption features} \label{section:telluric_absorption_features}
The Earth's annual motion around the solar system barycentre induces a time-dependent shift in the observed radial velocity of any given star.
In the stellar rest frame, telluric absorption features therefore appear to change their wavelengths with orbital phase.
A telluric feature which is blended to differing extents in different measurements of the same stellar feature causes a difficult-to-account-for scatter in the stellar feature's measured wavelength.
Even extremely weak telluric features can cause problematic shifts in the measured wavelength of stellar features at the  \SI{<10}{\meter\per\second} level of precision required for this work.
As our spectra were chosen to have SNR \(\geq200\), any background features with a depth greater than $\sim$0.1\% of the continuum are distinguishable from noise.

The centroid of an absorption feature blended with a weaker feature is approximately the weighted mean of the wavelengths of the two lines, weighted by their normalized depths.
The velocity shift of the centroid $\Delta v_c$ from the blended lines is then \citep[Equation 2,][]{Murphy2007a}
\begin{equation} \label{equation:feature_velocity_shift}
    \Delta v_{\mathrm{c}} 
    \approx \Delta v_{\mathrm{sep}} \frac{D_2 / D_1}{1+D_2 / D_1},
\end{equation}
where $D_1$ and $D_2$ are the depths of the two features, and $\Delta v_\text{sep}$ is the velocity separation between them.
At HARPS's resolution of \SI{2.6}{\kilo\meter\per\second}, a feature with normalized depth 0.1 just barely blended with another feature with depth $D_1/D_2=100$ times weaker (0.1\% of the continuum) would have $\Delta v_c\approx \SI{26}{\meter\per\second}$.
We therefore chose to avoid telluric features 0.1\% of the continuum or stronger, so this represents a theoretical upper limit on our error from telluric feature blending.

To identify telluric features, we used a synthetic spectrum from the Transmissions Atmosph\'eriques Personnalis\'ees Pour l'AStronomie online service\footnote{\url{http://ether.ipsl.jussieu.fr/tapas/}} \citep[TAPAS,][]{Bertaux2014}.
This synthetic spectrum was convolved with a Gaussian kernel to the resolution of HARPS, and took into account contributions from Rayleigh extinction, \BPChem{H\_2O}, \BPChem{O\_3}, \BPChem{O\_2}, \BPChem{CO\_2}, \BPChem{CH\_4}, and \BPChem{N\_2O}.
The spectrum was computed at zenith angle (i.e., an airmass of 1) using atmospheric parameters representative of La Silla Observatory.
To find telluric features we used a moving window, comparing the median value of a 5-pixel window at each wavelength with the median value of a 50-pixel window.
We flagged wavelengths where the difference was more than 0.1\% as being part of a telluric feature.
Visual inspection of the results confirmed that this approach was successful in flagging regions of significant absorption compared to the local continuum.

The maximum peak-to-peak amplitude of the change in the Earth's heliocentric radial velocity is \SI{\sim60}{\kilo\meter\per\second}, so at the resolution of HARPS any spectral feature within $\sim$\SI{30}{\kilo\meter\per\second} of a telluric feature can potentially be blended with it.
In addition to this annual change in radial velocity, the stars in our sample have their own peculiar radial velocities.
Avoiding telluric features thus requires considering the sum of the radial velocities of the Earth and each star.
On top of the annual \SI{\pm30}{\kilo\meter\per\second} range we added an additional \SI{\pm70}{\kilo\meter\per\second} range for the allowed stellar peculiar radial velocities, as this encompassed all 130 stars in the sample.
We then applied the resulting mask of \SI{\pm100}{\kilo\meter\per\second} to each flagged wavelength, and took the union of all such masks to create an overall set of masked regions of the spectrum to avoid selecting transitions in.
After these steps we were left with 783 possible transitions from the initial 8843.

\subsubsection{Blending with stellar absorption features} \label{section:transition_blendedness}

\begin{figure*}
  \begin{center}
    \includegraphics[width=\linewidth]{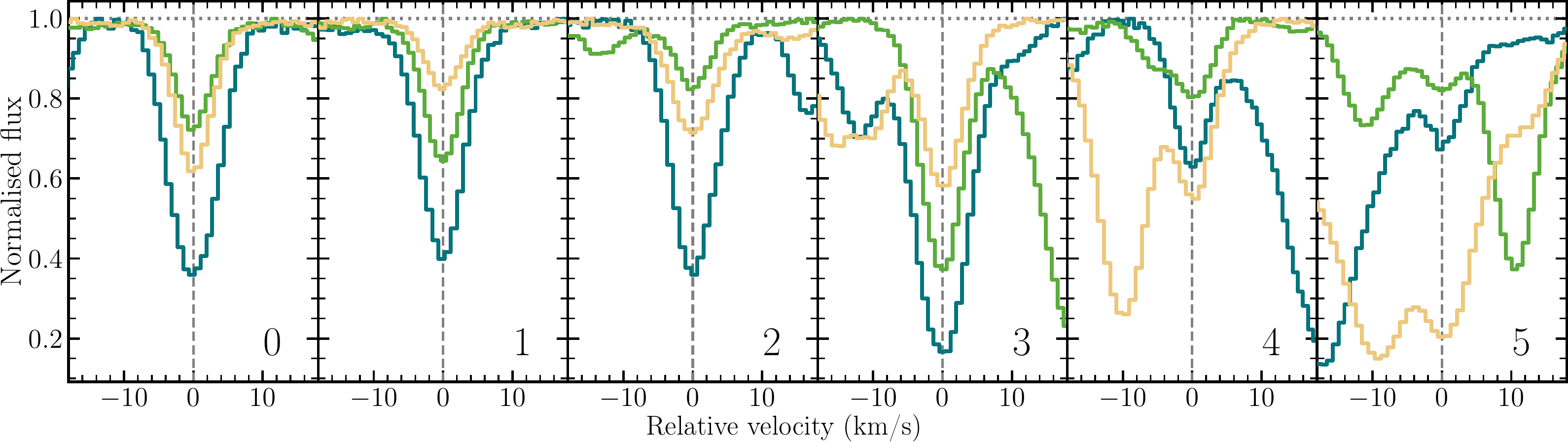}
    \caption{
    Normalised flux in absorption features showing examples of different ``blendedness'' values, from 0 on the left (no blending apparent) to 5 on the right (extremely blended).
    Blendedness is indicated by the number in the lower-right of each panel.
    Each panel shows three transitions chosen from across a wide spectral range ($\sim$4200--\SI{6250}{\angstrom}) and with a variety of depths, with the feature centroid marked by the vertical dashed line.
    At this SNR the flux uncertainties are smaller than the thickness of the lines.
    }
    \label{figure:blendedness_examples}
  \end{center}
\end{figure*}

In addition to time-variable blending with telluric features, blending with other stellar features is a concern.
Features showing greater blending have larger star-to-star systematic variance on average, as we demonstrate in \ref{section:sys_err:blendedness}.
To mitigate the effects of blending, we visually inspected each transition in the same high-SNR HARPS spectrum of the Sun mentioned above and rated each feature's ``blendedness'' on a scale of 0 (no blending apparent) to 5 (extremely blended).
\ref{figure:blendedness_examples} provides some representative examples of absorption features from each blendedness category.
Based on inspecting many examples from each category, we conservatively decided to use only transitions with blendedness values of $\leq2$.
This left us with a final sample of 164 transitions, out of 785 from the previous step. 
We test the effects of blendedness on the residual scatter in pair measurements across stars in \ref{section:sys_err:blendedness}, which showed that pairs with at least one transition with a blendedness of 4 or 5 had significant additional scatter in their separations, beyond what was accounted for by other uncertainties.

\subsection{Selection of transition pairs} \label{section:transition_pair_selection}
Absorption features of different normalised depths form over different ranges in stellar photospheres.
Conditions changing throughout the photosphere then imprint different asymmetries on these features \citep{Dravins1982, Dravins2008}.
Features of similar depths, however, should be affected by similar conditions and have similar asymmetries.
To minimise systematic errors from this source, we required the normalised depths of both transitions in each pair to be within 0.2 of each other.
We test this assumption in \ref{section:feature_depth_differences} and find no detectable systematics with this limit.

HARPS is known to have intra-order distortions in its calibration scale \citep{Molaro2013}, so to minimise them we set an upper limit of \SI{800}{\kilo\meter\per\second} on pair separations.
A lower limit of \SI{9.1}{\kilo\meter\per\second} was also imposed to avoid transitions blended at HARPS's resolution.
We placed no restrictions on element, ionization state, or blendedness for either transition (other than both transitions having blendeness\,$\leq2$).
The final pair sample is comprised of 229 pairs of transitions whose separations are to be compared between stars.
Of these pairs, we found that 54 could be measured simultaneously on two adjacent echelle orders (i.e., both transitions in each pair fell in an overlapping spectral region on both diffraction orders and could be measured in each).
We treat these `instances' of such pairs as separate measurements, which allowed us to perform the consistency checks on the HARPS CCD reported in \ref{section:cross_order_instances}.

\section{Details of analysis procedure} \label{section:analysis}
\subsection{Overview of HARPS} \label{section:HARPS}
For this work we used the High-Accuracy Radial velocity Planetary Searcher (HARPS) spectrograph, located on the European Southern Observatory's (ESO) \SI{3.6}{\meter} telescope at La Silla Observatory, Chile \citep{Pepe2002}.
HARPS is a high resolution ($\mathrm{R}\approx115000$), fibre-fed, cross-dispersed echelle spectrograph covering the visible portion of the spectrum from 380 to \SI{690}{\nano\meter} \citep{Mayor2003}.
The HARPS detector is composed of two 2048$\times$4096 15$\times$15\,\si{\micro\meter} pixel CCDs, each of which is composed of sixteen 1024$\times$512 pixel segments from the photolithography stepping process \citep{ESO2001}.
These segments are arranged in two rows stacked in the cross-dispersion direction ($y$) and eight in the dispersion direction ($x$)  for each CCD.
Echelle diffraction orders from 89 to 114 fall on the ``red'' CCD, while orders 116 to 161 fall on the ``blue'' CCD, with order 115 falling in the gap between them and not observed \citep{ESO2019}.
The CCD arrangement and characteristics are important to consider for this analysis, hence why we specify them here.

HARPS is encased in a vacuum chamber ($\mathrm{pressure}<\SI{0.01}{\milli\bar}$) which is kept temperature stabilized with an expected long-term stability of \SI{0.01}{\degreeCelsius} in order to minimise instrumental drifts \citep{Mayor2003}.
Despite this, HARPS still experiences wavelength calibration shifts on both nightly and longer-term scales.
To combat nightly shifts, HARPS uses a dual-fibre design which allows for spectroscopy of two different sources simultaneously on adjacent parts of the detector \citep{Mayor2003}.
Prior to each night's observations, several calibration images are taken with light from a thorium-argon (ThAr) lamp illuminating both fibres to establish an absolute calibration scale.
During the night the stellar target is acquired using the A fibre and a secondary calibration source is simultaneously acquired with the B fibre, allowing for tracking of shifts over the course of the night.
The secondary source was a ThAr lamp prior to 2012, with a Fabry-Perot (FP) etalon being used increasingly often since.

\subsubsection{Change of optical fibres used in HARPS} \label{section:harps_fiber_change}
HARPS was constructed with optical fibres (which convey light from the telescope to the instrument) with a circular cross-section.
Due to improvements in fibre design and better knowledge of the systematic effects imparted by circular fibres, between 19 May--3 June 2015 HARPS was upgraded with new, octagonal fibres \citep{LoCurto2015}.
This upgrade caused a change in the instrumental profile of the spectrograph, revealed by changes in the measured wavelengths of individual features of up to \SI{\sim100}{\meter\per\second} \citep{LoCurto2015, Dumusque2018}.
We have therefore kept observations from before and after the fibre change separate, and only compared measurements taken in the same era with each other.
These two eras are referred to as the `pre' and `post' fibre change eras to distinguish them.

\subsection{Creating uncertainty arrays} \label{section:HARPS_uncertainties}
HARPS has an automated reduction pipeline called the Data Reduction System \citep[DRS,][]{Rupprecht2004}, which automatically extracts and combines echelle orders in an observation into a final 1D spectrum.
However, these 1D spectra do not include uncertainty arrays, which are necessary for analysing individual absorption features.
They have also been corrected for the echelle orders' blaze functions, which removes crucial information on original photon-counts-per-pixel that could otherwise be used to create uncertainty arrays.
\citet{Dumusque2018} created uncertainty arrays for HARPS spectra by using intermediate data products from the DRS, in the form of extracted 2-dimensional spectra (``e2ds'') files containing the extracted flux from each order prior to any additional processing.
We followed this procedure, creating uncertainty arrays by taking the square root of the quadrature sum of the photon count per pixel plus HARPS's dark and read-out noise as given in \citet{Dumusque2018}.
We then used the nightly blaze correction functions available from the HARPS archive\footnote{\url{http://archive.eso.org/cms/eso-data/eso-data-direct-retrieval.html}} to correct the flux and error arrays.

\subsection{Wavelength calibration} \label{section:HARPS_calibration}
Wavelength calibration for HARPS is performed by fitting a third-degree polynomial to the dispersion relation between laboratory wavelengths of lines from a ThAr reference and their positions on the CCD.
A separate polynomial is fit for each extracted echelle order in the e2ds files, and the coefficients of these 72 polynomials are stored in the FITS headers of the file.
The wavelength scale can then be reconstructed using Equation (1) from \citet{Dumusque2018}.
However, there are complications in the process, which we detail below, which must be taken into consideration for high-precision work with individual transitions.

\subsubsection{Variation in CCD pixel sizes} \label{section:pixel_column_width}
HARPS's wavelength scale is known to have distortions of up to \SI{\sim100}{\meter\per\second} across the free spectral range of echelle orders \citep{Wilken2010a,Molaro2013,Bauer2015}.
Using a laser frequency comb (LFC) with calibration precision of a few \si{\centi\meter\per\second}, \citet{Wilken2010a} found discontinuities due to pixel column width variations at intervals of 512 pixels, corresponding to edges of sub-divisions of the CCD from the photolithography stepping process.
\citet{Coffinet2019} were able to measure the pixel column width variations using flat-field frames, and demonstrated improved accuracy of the wavelength scale by including their effects in the calibration process.
They also re-calibrated many of the nightly ThAr exposures in the HARPS data archive to account for this effect, generating polynomial coefficients for the new wavelength solutions.
We received these updated coefficients directly from the authors, though they were not available for all the nights on which our archival observations were taken (C. Lovis, priv. comm.).

\begin{figure}
  \begin{center}
    \includegraphics[width=\linewidth]{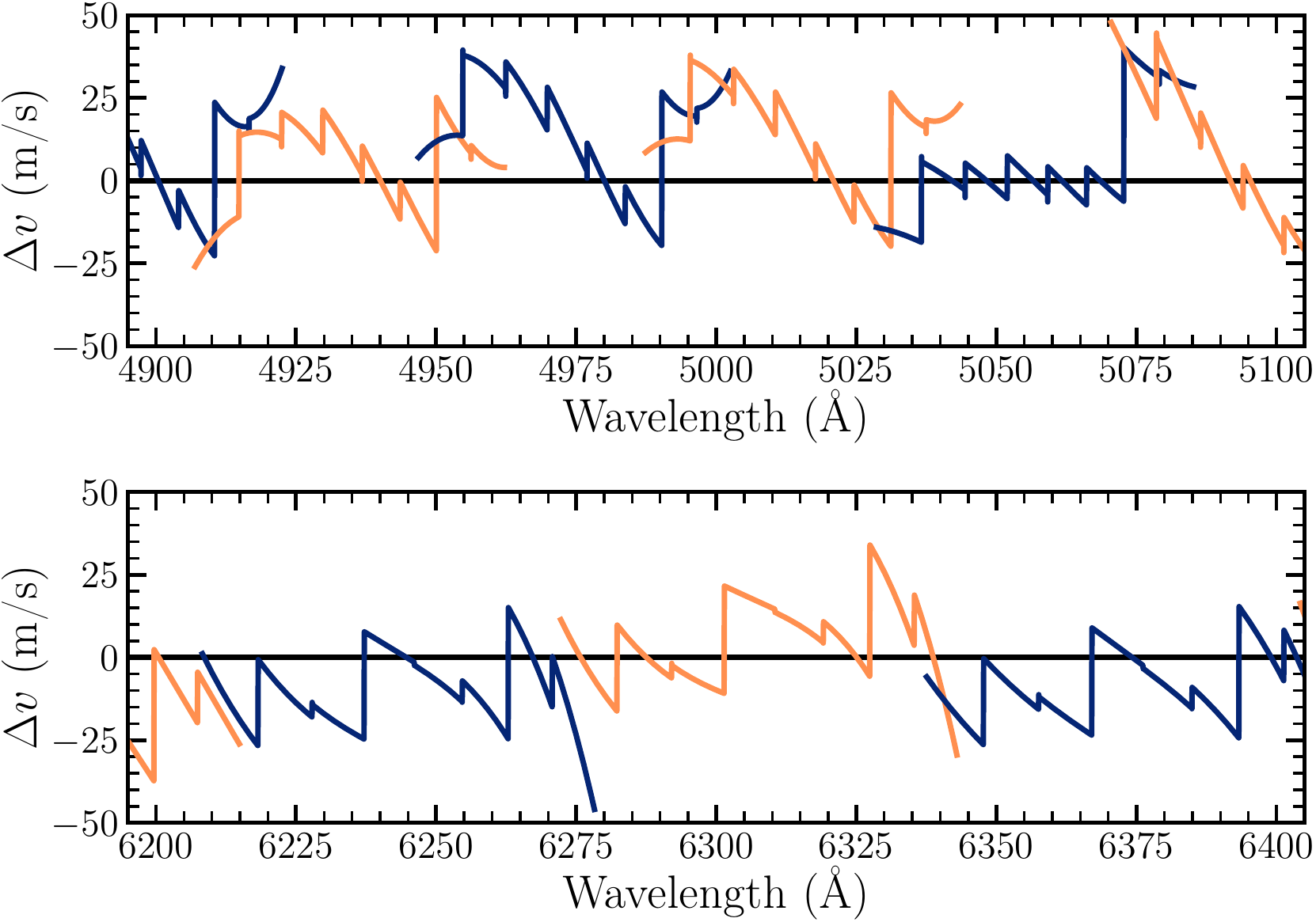}
    \caption{
    The difference, in velocity space, between the old and new HARPS wavelength solutions, plotted as a function of wavelength for selected portions of HARPS's spectral range.
    The zero line represents the calibration by laser frequency comb, while the jagged lines represent the distortion that would occur if not using the new solution.
    Alternate orders are plotted with contrasting colors to distinguish them.
    The top panel shows a selection from the blue CCD, while the bottom shows a selection from the red CCD.
    The discontinuities represent the effects of irregular pixel column widths every 512 pixels on the CCD, as described in \ref{section:pixel_column_width}.
    }
    \label{figure:old_new_wavelength_scales_comparison}
  \end{center}
\end{figure}

\ref{figure:old_new_wavelength_scales_comparison} shows the difference between the old and new wavelength scales for representative portions of HARPS's spectral range on both of its CCDs.
The discontinuities at 512-pixel intervals are clearly apparent.
Differences at the level of $\SI{\sim50}{\meter\per\second}$ represent distortions that would be present in our results if not using the updated calibrations.
Given the significant improvement in the accuracy of the wavelength scale demonstrated in \citet{Coffinet2019}, we used only observations taken on nights for which we had new calibration coefficients.

\subsubsection{Choice of wavelength calibration function} \label{section:harps_calibration_function}
Building on the results from \citet{Coffinet2019}, \citet{Milakovic2020} used LFC calibration to search for additional systematic effects in HARPS's wavelength scale.
They found that the choice of polynomial used to fit the dispersion relation mattered significantly: fitting each echelle order with a single polynomial (as in the DRS) left distortions in the wavelength scale, even after correcting for the pixel width variations.
These remaining distortions had amplitudes of up to \SI{10}{\meter\per\second}, and varied on length scales of tens to hundreds of \si{\kilo\meter\per\second}, compared to the average width of \SI{\sim3300}{\kilo\meter\per\second} of an echelle order.
However, fitting each echelle order piece-wise with a separate polynomial for each 512-pixel segment effectively removed these residual distortions.
A map of the residuals between this improved wavelength scale based on the LFC and the ThAr-calibrated HARPS wavelength scale was provided to us (D.\ Milakovi\'c, priv. comm.).
We used this map to correct the measured wavelength of each spectral feature in our analysis based on its recorded position on the CCD.

The residual map only covered three out of four of the vertical 1048-pixel `blocks' on the detector (both blocks on the red CCD, and one on the blue CCD) due to wavelength coverage limitations of the LFC.
The block not covered by the LFC contains echelle orders 135--161, where 22 of the 164 transitions in the sample are located (the reddest being at \SI{4589.484}{\angstrom}).
Milakovi\'c et al. found that the residuals did not vary significantly between individual orders in the same block, and thus used a single, averaged, set of residuals per block.
Visually, plots of the residuals for all three blocks look quite similar in form, so for correcting the fourth block we used the residuals from the other block on the blue CCD.
Given the level of agreement between the three sets of residuals, any resulting errors from using residuals from another block should be smaller than those from not instituting any correction, at the level of a few \si{\meter\per\second}.

\subsection{Measuring pair separations} \label{section:measuring_pair_separations}
Due to the large number of features to be fit in each spectrum, and in order to remove human bias as much as possible, we used an automated absorption feature fitting routine.
This fitting process is described in section 4.1 of \citetalias{Berke2022a}, but, to summarize, we used a custom Python package called \textsc{VarConLib}\footnote{Short for ``Varying Constants Library,'' code available at \url{https://github.com/DBerke/varconlib}.} to fit a four-parameter integrated Gaussian function to the central seven pixels of each feature, and took its centroid as the wavelength of the transition.
Measuring pair separations rather than comparisons to absolute wavelengths should greatly reduce any systematic effects from fitting asymmetric absorption features with a symmetric function, while fitting only the feature cores also reduces any potential impact \citep{Dravins2008}.
The amplitude of the Gaussian model was not constrained, so any cases where it was positive were discarded as this indicated that the feature was weak enough that a noise peak had been fitted instead (as sometimes happened to weak features in low-metallicity or high-temperature stars).
The wavelength scale corrections from \ref{section:harps_calibration_function} were applied to each wavelength,
and the final velocity separation for each pair was taken as the difference between the two measured wavelengths of its component transitions: $\Delta v_\text{pair}=2c(\lambda_\mathrm{red}-\lambda_\mathrm{blue})/(\lambda_\mathrm{red}+\lambda_\mathrm{blue})$, where $\lambda_\mathrm{red}$ and $\lambda_\mathrm{blue}$ refer to the wavelengths of the red and blue transitions in each pair.

\subsection{Modelling pair separation change as a function of stellar parameters and rejecting outliers} \label{section:pair_separation_variation}
An important assumption in the solar twins method is that we are able to compare pair separations in stars that are very similar.
It is important to check this assumption and correct for changes in separation as a function of important atmospheric parameters such as \Teff, [Fe/H], and $\log{g}$.
Such changes can then be modelled as a function of those parameters within a defined range wherein stars can be compared.
This is especially important for future work building on our results by searching for distant solar twins several \si{kpc} closer to the Galactic Centre, as described in accompanying papers (\citealt{Lehmann2022},  Liu et al. in prep., Lehmann et al. in prep.).
Uncertainties in stellar parameters of distant potential solar twins discovered in target finding campaigns will likely be large, on the scale of the range defined as solar twins.
The solar twins method must therefore be usable over a wider range, such as that defined by solar analogues in \ref{equation:star_definitions}, to increase the potential sample size.

\begin{figure}
  \begin{center}
    \includegraphics[width=\linewidth]{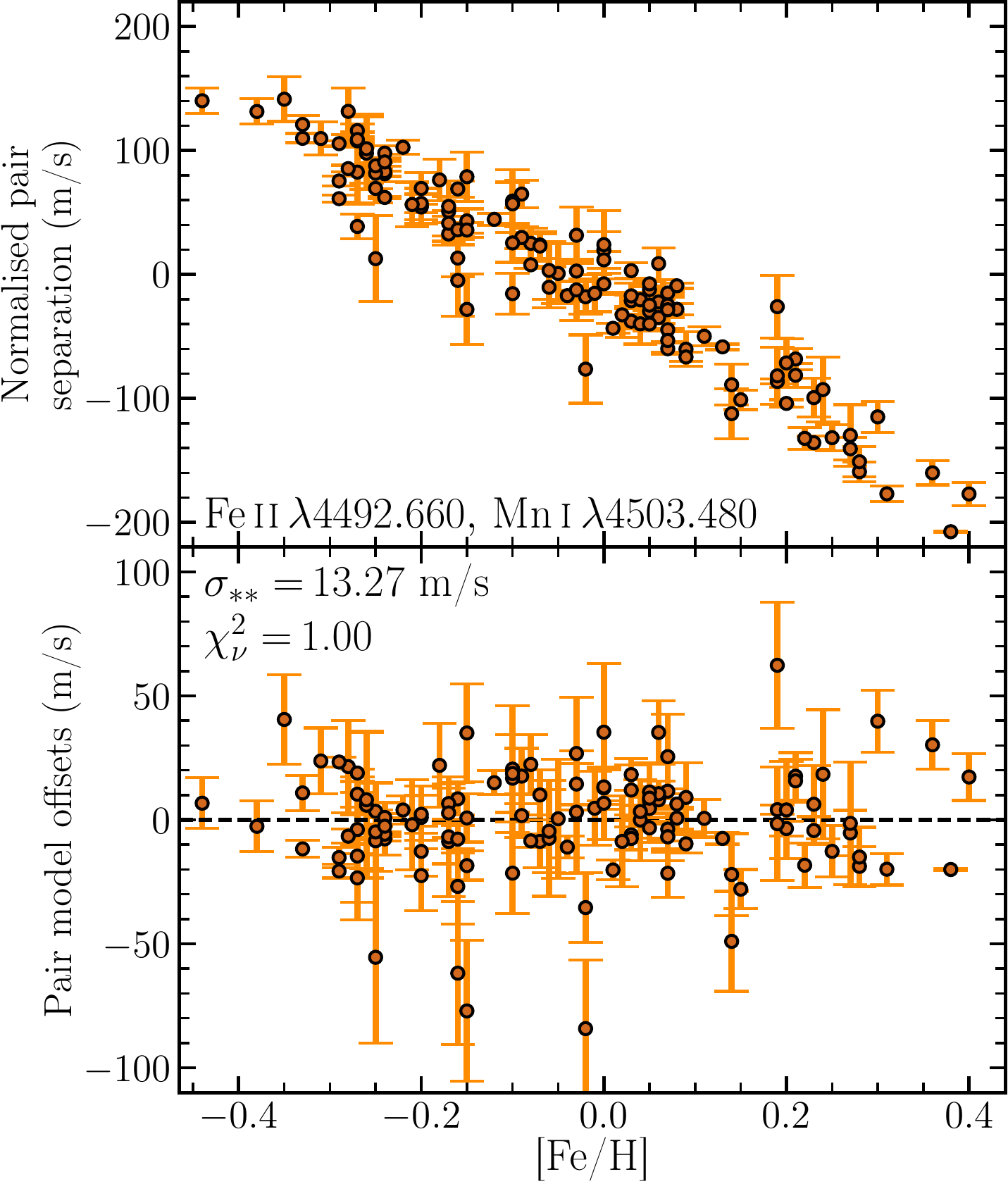}
    \caption{The weighted mean velocity separation (per star) between the pair of transitions $\ion{Fe}{ii}\,4492.660$ and $\ion{Mn}{i}\,4503.480$ as a function of stellar metallicity.
    The top panel shows the spread in the measured separations after shifting the distribution to a mean of zero.
    The bottom panel shows the residuals after fitting the values in the top panel as described in \ref{section:pair_separation_variation}, with a $\chi^2_\nu$ value of 1.00. Only the statistical error is shown in both panels, without the \sigsys value found for this pair (explained in text) added in quadrature, though it is given in the lower panel.
    Note the different vertical scales between the two panels.}
    \label{figure:pair_separation_change}
  \end{center}
\end{figure}

Initial investigation of pair separation measurements showed that the majority of pair separations varied clearly and systematically with one or more of the stellar atmospheric parameters \Teff, [Fe/H], or $\log{g}$.
The top panel of \ref{figure:pair_separation_change} shows a representative pair where the separation changes by nearly \SI{400}{\meter\per\second} between $-0.45$ and 0.45 dex in metallicity.
The magnitude of the change in this pair is on the high end of the range seen across our pair sample, but the majority of pairs exhibit a similar change as a function of one of the parameters listed above.

To model the change in pair separation with atmospheric parameters, we performed a multivariate fit of the (inverse-variance) weighted mean, across all observations of each star, as a function of \Teff, [Fe/H], and $\log{g}$.
%For each star we used the weighted mean value (using inverse-variance weighting) of the pair separation across all observations of the star.
We tested various functions, and found that even a function linear in all three variables fit the data adequately, as measured by the reduced chi-squared value (per degree of freedom) $\chi^2_\nu$ of the residuals.
A quadratic function did slightly better, but higher-order functions or functions with cross-terms produced no statistically-significant improvement.
We therefore used the following quadratic function:
\begin{multline} \label{equation:fitting_function}
    \Delta v_\mathrm{pair} = a + \sum_{n=1}^{2} b_n \Teff^n + c_n \textrm{[Fe/H]}^n + d_n (\log g)^n,
\end{multline}
where $\Delta v_\mathrm{pair}$ is the pair separation value and $a, b_n, c_n$, and $d_n$ are coefficients to be determined from the fit for each pair.
The bottom panel in \ref{figure:pair_separation_change} shows the residuals left after subtracting the appropriate model value from the weighted mean of each star.
These residuals, which we term `pair model offsets,' show no signs of remaining structure, indicating that the quadratic model is an adequate fit to this pair.
Similar residual plots for the same pair versus \Teff and $\log{g}$ show the same flat distribution without discernible structure.
We visually inspected plots of the pair model offsets for each parameter for every pair and found the same general results.
We interpret this as indicating that the fitting process successfully removed any systematic low-order variation in $\Delta v_\textrm{sep}$ to below the level of the remaining scatter.

Importantly, nearly all pair model offsets still showed greater scatter than expected from the statistical uncertainties.
For an individual star, this represents a systematic deviation from the pair separation model and we discuss potential causes of such deviations in \ref{section:elemental_abundances_and_isotope_ratios}.
To measure this additional star-to-star scatter, we introduced an additional systematic term, denoted \sigsys.
We used the following iterative $\sigma$-clipping procedure to model each pair while simultaneously determining \sigsys and identifying outliers at each step.

In each iteration the pair separations are fitted using \ref{equation:fitting_function}.
The uncertainty for each star is then set to the quadrature sum of its statistical error on the weighted mean and \sigsys (set to zero in the first iteration).
Any outliers greater than $4\sigma$ are then flagged.
In the second iteration, if $\chi^2_\nu<1$ for the distribution then there is no additional scatter beyond that expected from the errors and iterations end.
If $\chi^2_\nu>1$ for the distribution, \sigsys is initially defined as $\sqrt{\chi^2_\nu-1}$ times the median uncertainty of the distribution and iterations continue.
At each step after the second, a new value of \sigsys is estimated according to \(\sigma_{**,n+1} = \sigma_{**,n}\cdot(\chi^2_\nu)^{2/3}\) (the power must be $<1$, but its exact value merely sets how fast the iteration converges to a solution; 2/3 was chosen as a good compromise after some minor experimentation, with the vast majority of pairs taking $<25$ steps to converge to a solution).
Iterations proceed until $\chi^2_\nu$ is within 0.001 of unity and either no stars were flagged as outliers in that step or the same stars were flagged as in the previous step.
Once iterations have finished, \sigsys and the best-fit parameters for \ref{equation:fitting_function} for the pair are saved.
Any outliers $>4\sigma$ in the final iteration step are flagged, and not used in further analysis.

\subsubsection{Change in star-to-star scatter with stellar parameters} \label{section:stellar_parameter_dependence}

\begin{figure*}
  \begin{center}
    \includegraphics[width=\linewidth]{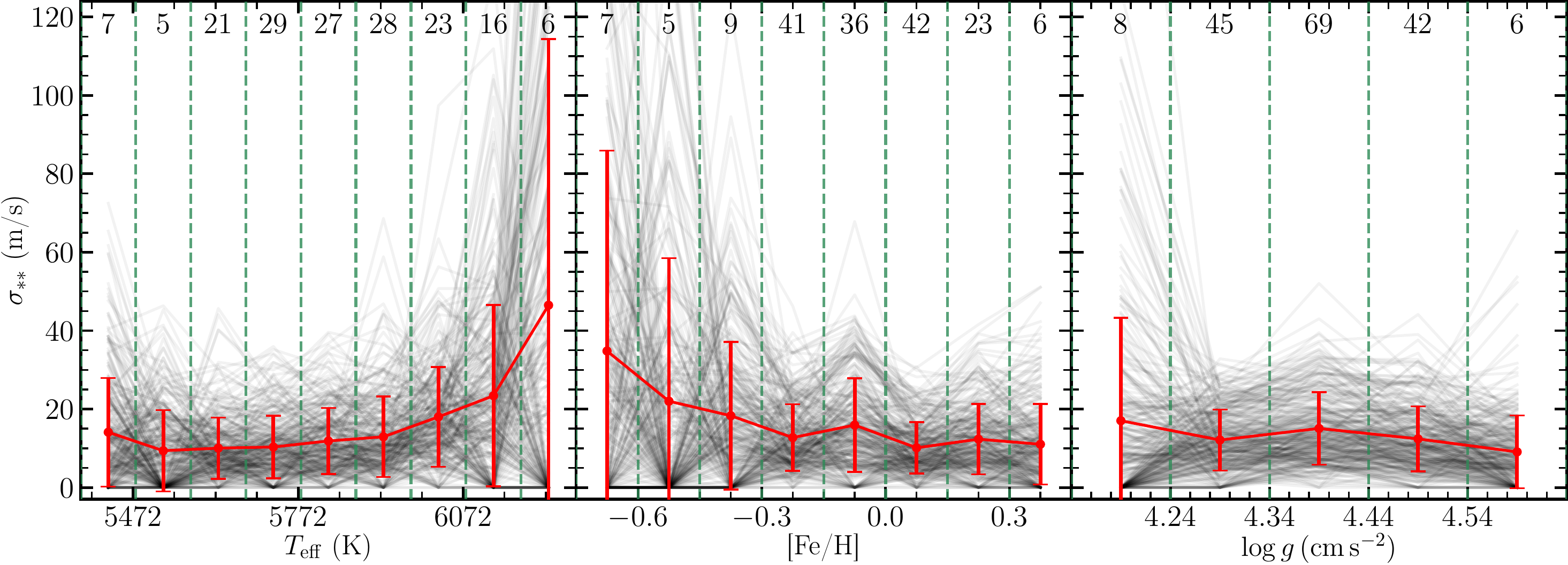}
    \caption{
    Search for variation in the star-to-star scatter as a function of stellar parameters.
    In each panel the grey lines show the value of \sigsys for individual pairs within each bin denoted by the vertical dashed lines.
    The mean and RMS of the \sigsys values in each bin are plotted as the red line with error bars.
    The number of stars in each bin is listed along the top of each plot.
    The vertical scale is set such that individual black lines may exceed the height of the plot, as it is primarily the aggregate behaviour of all pairs that is important.
    }
    \label{figure:stellar_parameter_dependence}
  \end{center}
\end{figure*}

If $\sigsys$ varies systematically with stellar parameters, using a single value of $\sigsys$ for each transition pair may under- or over-estimate the errors for individual stars.
To look for such variation, for each pair we split all the stars in the stellar sample into bins in \Teff, [Fe/H], and $\log{g}$ (9, 8, and 5 bins, respectively), then repeated the iterative procedure above on each bin.
The result was a set of \sigsys values for each pair which are plotted in \ref{figure:stellar_parameter_dependence} across the range of all three parameters.
For $\log{g}$ no significant trend is seen.
\Teff shows a slight trend of increasing mean and scatter in \sigsys values with increasing temperature, while [Fe/H] shows a similar trend with decreasing metallicity.
Both results can be explained as an effect of decreasing absorption feature depth: weaker features lead to a systematic increase in relative noise in the wavelength measurements, especially in features already weak at solar values.

\ref{figure:stellar_parameter_dependence} is an important result for this work, as it illustrates the range in which our assumption of `similarity' between stars -- which underlies the solar twins method -- is actually valid.
While the change in \sigsys with stellar parameters could potentially be modelled and accounted for, we instead use \ref{figure:stellar_parameter_dependence} as a guide and revise the stellar sample accordingly.
Based on these results we use only stars with $\Teff\leq\SI{6072}{\kelvin}$ (\SI{300}{\kelvin} hotter than the solar value) and $\mathrm{[Fe/H]}\geq-0.45$, as the mean and scatter in \sigsys values increase -- potentially too much -- beyond those points.

\subsubsection{Modelling transition wavelength change as a function of stellar parameters and rejecting outliers} \label{section:transition_wavelength_variation}
Changes in pair velocity separations can arise from changes in one or both of the component transitions as a result of differences in stellar parameters.
To correct for such shifts in transitions and reject outlying wavelength measurements, we performed the same fitting procedure described in \ref{section:pair_separation_variation} on each transition prior to fitting pair separations.
The quantity fitted in this case was the velocity separation of the measured wavelength of the feature from its expected laboratory value, rather than the pair separation.
Here we encounter an issue not seen with pair separations -- by construction -- where stars hosting planets have radial velocities which vary with time, causing their transition separations to vary systematically between observations.

To correct for these time-varying radial velocities, we applied the following procedure to each star.
We started with the radial velocity given for the star stored in each observation's header, which generally had an accuracy of \SI{10}{\meter\per\second}.
Transitions were then fit in each observation of a star (for a complete description of the process see section 4 of \citetalias{Berke2022a}), and their velocity separations from their expected wavelengths measured.
These velocity separations were found to be distributed normally, so for the separations in each observation we subtracted their median to align all observations in the star's rest frame.
We note again that this step was unnecessary (and not performed) for the pair separations because of their differential nature, and the existence of these systematic effects demonstrates the need for such a differential approach for measuring any variation in $\alpha$ between stars, rather than direct comparison with laboratory wavelengths.
Having performed this correction, a more conservative $\sigma$-clipping threshold (compared to the pair separation fitting) of $3\sigma$ was used to more stringently identify outlying individual measurements (such as from cosmic ray hits, described in \ref{section:cosmic_rays}).
These outliers were not used in the subsequent pair separation analysis described in \ref{section:pair_separation_variation}.

We initially attempted to use the reduced $\chi^2$ statistic for individual feature fits to reject outlying measurements, but this proved unreliable.
No correlation was observed between $\chi^2_\nu$ for a feature fit and the statistical significance of its velocity separation from the expected wavelength.
Additionally, average $\chi^2_\nu$ values for the same transitions were found to vary between stars, making it impossible to assign a single cut-off suitable for all stars.
The infeasibility of using $\chi^2_\nu$ for outlier rejection in this case is expected because there are only three degrees of freedom (seven pixel fluxes minus four parameters) in the fitting model, and the $\chi^2$ distribution for three degrees of freedom is very broad.
We thus adopted the method reported above instead.

\section{Systematic errors} \label{section:systematic_errors}
In this section we detail all sources of systematic error considered and our estimates of how much they factor into the final systematic error budget.
A systematic error here refers to a systematic difference in a pair's separation between stars, as measured at a specific location on HARPS's CCD.
Many of the effects described here are therefore position-dependent, caused by pairs being measured at slightly different locations on the CCD.
Achromatic effects which affect the entire spectrum (such as Doppler shifts) are not a concern, except insofar as they shift the measurement location of pairs and potentially trigger position-dependent effects.

\subsection{Varying CCD measurement location} \label{section:ccd_position_systematic_errors}
The fixed format of HARPS's construction provides a great benefit for the solar twins method, as observations of the same transition will always fall in a small range around the same position on the CCD, limiting certain systematic errors.
This range is caused by two factors: annual modulation of the Earth's barycentric velocity, and the radial velocities of stars in the stellar sample.
In total, transitions are measured within a range of \SI{\pm100}{\kilo\meter\per\second}, of which \SI{\pm30}{\kilo\meter\per\second} comes from the Earth's orbital motion and the rest comes from the selection limits we imposed in \ref{section:telluric_absorption_features}.
Each transition is thus measured within a range of \SI{\pm125}{pixels} on the CCD (cf. a total range of \SI{4096}{pixels} in each echelle order). 

The instrumental profile of HARPS varies slightly across the CCD \citep{Zhao2014,Zhao2021}, so changes in measurement location cause a systematic variation in measured transition wavelengths \citep{Dumusque2018}.
While we have corrected known systematics in HARPS's wavelength calibration across echelle orders, as detailed in \ref{section:HARPS_calibration}, additional systematic errors related to varying instrumental profile or other sources may remain in our measured transition pair separations.
We detail two possible such systematics below.

\subsubsection{Variation in pair separation across the CCD} \label{section:cross_order_instances}

\begin{figure*}
  \begin{center}
    \includegraphics[width=\linewidth]{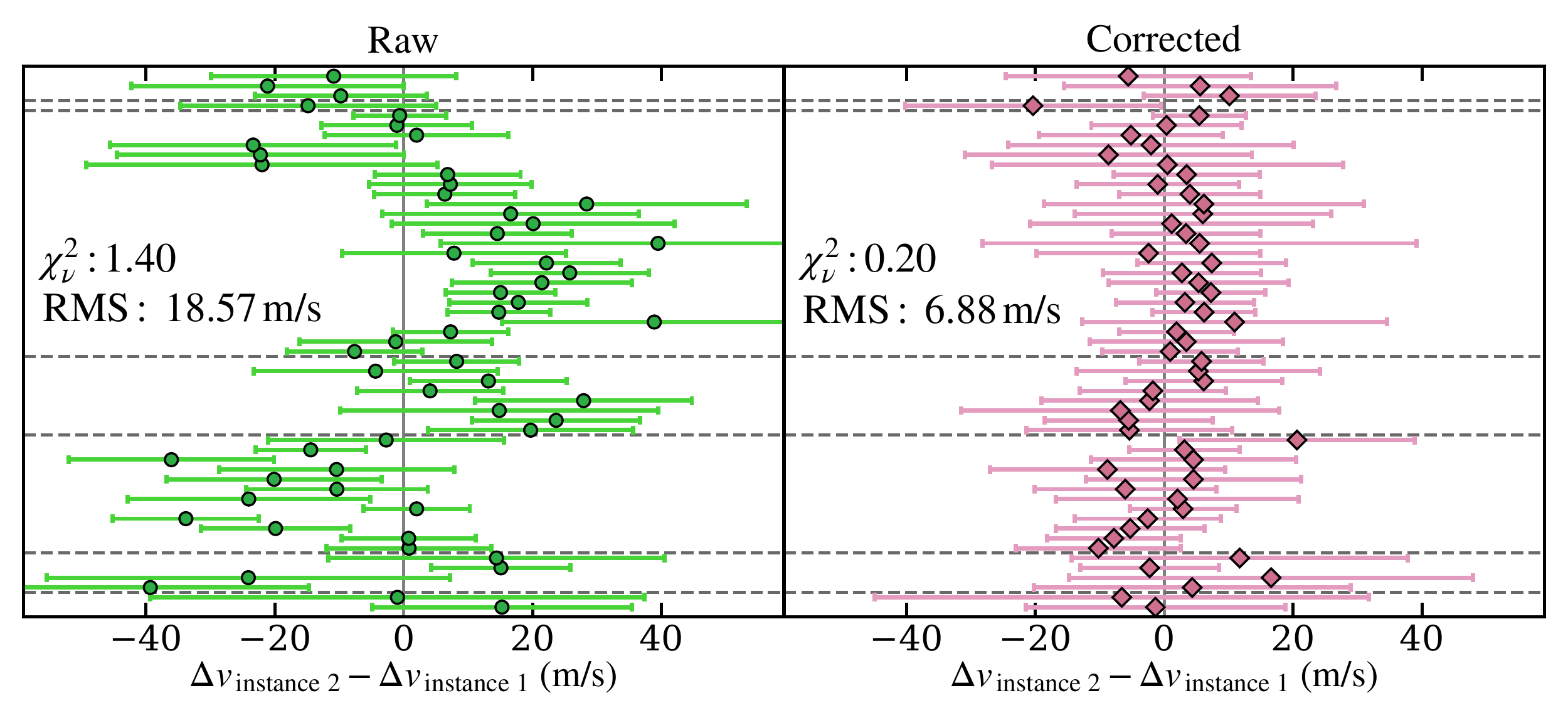}
    \caption{
    The difference in pair separation between both instances of 54 pairs which are measured on opposite ends of the CCD.
    Each point is the weighted mean value from 87 observations of the solar twin HD~146233.
    \textbf{Left panel:} Raw pair separations, as measured.
    \textbf{Right panel:} Separations after each instance has been corrected by its own best-fit model, i.e., pair model offsets.
    The dashed horizontal lines demarcate groups of pairs found in the same echelle orders, showing some correlation between pairs found in similar areas of the CCD.
    The $\chi^2_\nu$ and RMS values of the distributions plotted are provided as a guide only, as multiple pairs can (and do) share transitions and are thus not independent.
    }
    \label{figure:duplicate_pairs}
  \end{center}
\end{figure*}

HARPS is a grating cross-dispersed spectrograph where the width of the detector exceeds the free spectral range, which produces overlap in spectral coverage between adjacent echelle orders\footnote{The overlap is most pronounced at the blue end, decreasing to no overlap in the reddest two orders.}.
Due to this overlap 54 transition pairs in the sample appear simultaneously on opposite sides of the CCD in two adjacent echelle orders.
We refer to these duplicates as `instances' of a pair and differentiate them by the echelle order in which they appear.
We treat each instance independently in the modelling process described in \ref{section:pair_separation_variation}, enabling their use as probes to search for systematic errors across the width of the detector.

As duplicate instances are two measurements of the same pair, the difference in velocity separation between them should be consistent with zero if there are no systematic effects present.
The difference between each set of pairs is plotted using the weighted mean values from 87 observations of the solar twin HD~146233 in the left panel of \ref{figure:duplicate_pairs}.
Significant deviations from zero are visible in a number of pairs, with those in the same echelle orders often also showing correlated behaviour different from pairs in other orders.
We investigated the possibility of charge transfer inefficiency (CTI) causing these cross-CCD variations (\ref{section:charge_transfer_inefficiency}) but concluded that they are too large to be due to CTI.
They may be due to instrumental profile variations across the CCD, or remaining uncorrected systematic errors in calibration, but the use of a single star with many high-SNR ($\geq200$) observations indicates that the effect is instrumental rather than astrophysical in nature.

Duplicate instances can also be used as a consistency check of pair separations after correcting for known systematic errors.
The right panel of \ref{figure:duplicate_pairs} shows the difference between instances after correcting each by its own best-fit model as determined from \ref{section:pair_separation_variation}.
After this correction all but a single pair are consistent with zero difference at a $1\sigma$-level, and the scatter decreases by a factor of 2.6.
This plot demonstrates that significant systematic differences are present on opposite sides of the CCD, at up to a \SI{\sim20}{\meter\per\second} level, but also that our approach of modelling each pair independently is successful at removing them.
That further suggests an instrumental origin for these systematic effects.

These systematic offsets between opposite sides of the CCD also validate our initial decision to use the 2D spectra rather than the 1D data products from the HARPS DRS (as mentioned in \ref{section:HARPS_uncertainties}).
The merging of different spectral orders will introduce an unknown systematic error to any pair found within these overlapping spectral regions.
As 48 out of the 229 pairs used in this work are found in these overlap regions, we avoid the systematic errors that would occur in such 1D spectra.

\subsubsection{Variation from radial velocity changes} \label{section:radial_velocity_variation}
Changes in radial velocity between observations cause transitions to be observed in slightly different locations on the HARPS CCD, and as the previous section established, this has the potential to introduce systematic errors.
The Earth's barycentric velocity varies over the course of a year by \SI{\pm29.30}{\kilo\meter\per\second}, while the stars in the sample have a range of heliocentric radial velocities chosen to be between \SI{\pm70}{\kilo\meter\per\second}.
For the average spectral width of a HARPS pixel of \SI{0.825}{\kilo\meter\per\second}, this means that any transition in the sample could potentially be measured over a range of up to $\sim$\SI{250}{pixels} on the CCD\footnotemark.
\footnotetext{The shift due to the change in the Earth's barycentric radial velocity over the course of a single observation is no more than a few \si{\centi\meter\per\second} for observations in our sample, and is thus negligible for our purposes.}

\begin{figure}
  \begin{center}
    \includegraphics[width=\linewidth]{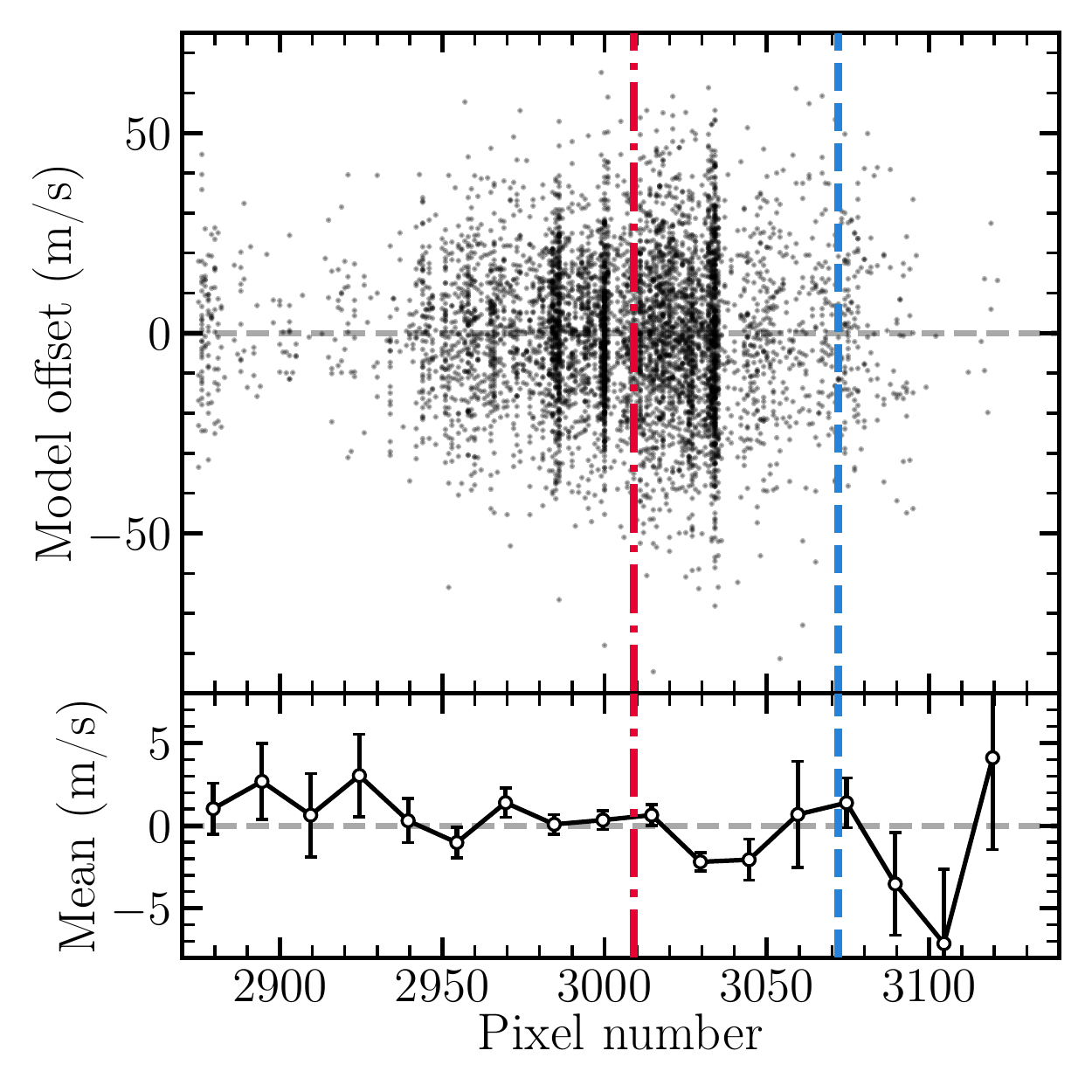}
    \caption{
    Variation of pair model offsets with different radial velocities of stars in the sample.
    \textbf{Upper panel:} Black points show the pair model offsets for all measurements of the pair \ion{Fe}{I}\,6138.313 -- \ion{Fe}{I}\,6139.390 across all stars.
    The horizontal axis denotes the pixel across the CCD at which the blue transition of the pair was recorded.
    \textbf{Lower panel:} The black line with errors bars shows the mean and its 1-$\sigma$ uncertainty for points in the upper panel in bins 15 pixels wide.
    The dashed-dotted (red) and dashed (blue) lines show the pixel at which the red and blue transitions crossed a boundary on the CCD between subsections.
    }
    \label{figure:radial_velocity_effects}
  \end{center}
\end{figure}

We searched for systematic variation in pair separation on this smaller scale of $\sim250$ pixels by considering all measurements of a pair whose transitions are spread across a CCD sub-boundary so as to probe the most likely source of systematic error in this scale.
\ref{figure:radial_velocity_effects} shows an example, for the pair \ion{Fe}{I}\,6138.313 -- \ion{Fe}{I}\,6139.390, where the upper panel shows the pair model offset plotted against the\ pixel on the CCD (in the dispersion direction) where the measurement was made.
No significant change in the mean (plotted for 15-pixel bins in the lower panel) is observed when either transition crosses the CCD sub-boundary, as indicated by the vertical lines.
At most, changes at the \SI{\sim5}{\meter\per\second} level may be seen, but there appears no evidence for significant changes.
Residual errors in wavelength calibration would be most expected in transitions crossing such a boundary, so their absence here supports a lack of significant systematic effects remaining from changing measurement location.
We checked three additional pairs where both transitions cross a CCD boundary, and found a similar lack of systematic variation with pixel in all of them (all four pairs crossed different boundaries).
This suggests that for other pairs where one or no transitions cross CCD boundaries any change as a function of radial velocity/pixel should be negligible.

\subsection{Stellar absorption feature blending} \label{section:sys_err:blendedness}

\begin{figure*}
  \begin{center}
    \includegraphics[width=\linewidth]{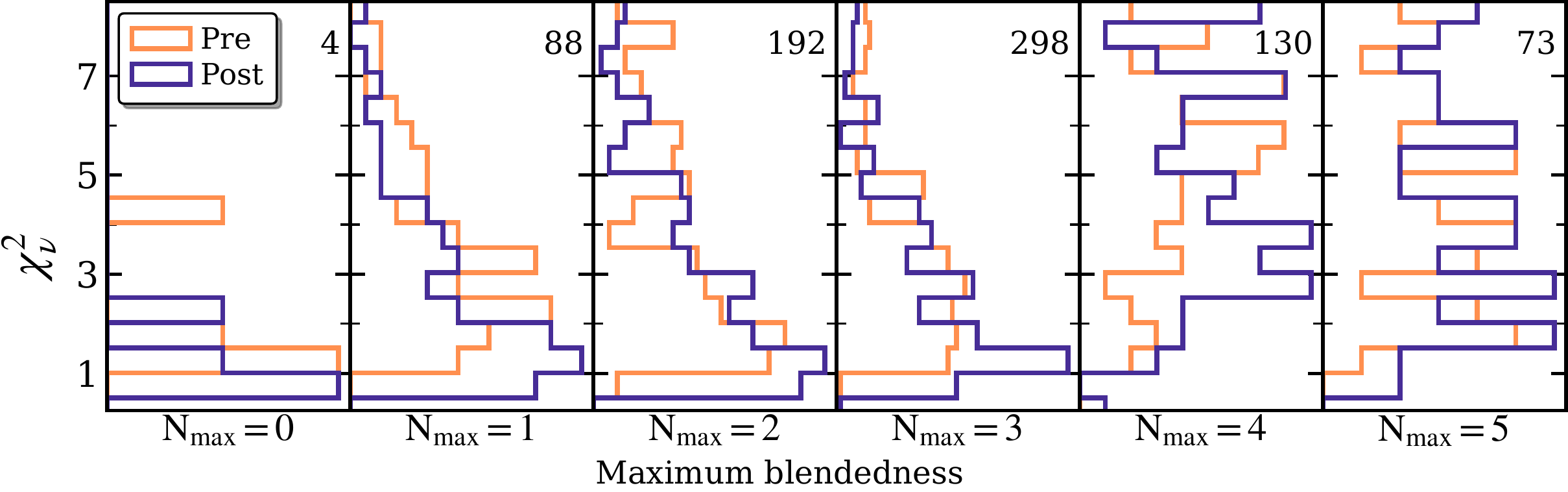}
    \caption{
    Effects of blending on residual scatter in pair separations between stars.
    Each panel shows, for pre- and post-fibre change data, the distribution of $\chi^2_\nu$ for the weighted mean pair separation of pairs with the maximum blendedness N$_\text{max}$ listed below the panel.
    The number of pairs in each category is given by the number in the upper-right of each panel.
    Below a maximum blendedness of 4, pair $\chi^2_\nu$ values tend to cluster closer to 1, while above it the distributions are less centered and more evenly scattered.
    For nearly all panels the distributions continue above the top of the plot; the focus of this plot is the behaviour close to unity (meaning no extra star-to-star scatter seen), rather than on the full range of each distribution.
    }
    \label{figure:pair_blendedness}
  \end{center}
\end{figure*}

As discussed in \ref{section:transition_blendedness}, each transition in the sample is assigned a ``blendedness'' value from 0 to 5.
The separation of a transition pair will be affected by the blending of its transitions, which will shift their centroids according to \ref{equation:feature_velocity_shift}.
These shifts will be stable across observations of the same star, and simply incorporated into the measurement of the pair's separation, where they can be modelled as in \ref{section:pair_separation_variation}.
However, even after this process there may be residual differences from star to star, caused by e.g. elemental abundance differences (we discuss this and other possibilities further in \ref{section:elemental_abundances_and_isotope_ratios}).
We would therefore expect that pairs with more highly blended transitions would show greater residual scatter between stars even after modelling out the variation across the stellar sample as described in \ref{section:pair_separation_variation}.

We sorted the pair sample into categories from 0 to 5 based on N$_\text{max}$ -- the blendedness of the most blended transition in the pair -- on the assumption that any scatter from the more-blended transition would dominate scatter from the less-blended one.
For each pair, $\chi^2_\nu$ was calculated for the distribution of its model-corrected, weighted mean separation across stars, without adding in the empirically-derived \sigsys value which would bring $\chi^2_\nu$ down to unity (\ref{section:pair_separation_variation}).
The $\chi^2_\nu$ distributions for the six blendedness categories (total 785 pairs) are shown in \ref{figure:pair_blendedness}, with results for the two observing eras shown separately.
The distributions for both eras look similar, as we would expect, though we note in passing that the $\chi^2_\nu$ distributions for the post-fibre change era tend to skew slightly closer to 1.
This is simply a different way of visualising  the decrease in \sigsys observed after the fiber change (and discussed in \ref{section:elemental_abundances_and_isotope_ratios}), which we attribute -- at least in part -- to a general improvement in HARPS's calibration accuracy \citep{LoCurto2015}.
The values in most of the panels of \ref{figure:pair_blendedness} extend past the top of the plot, as the focus is not on the overall range of the distributions but rather their behaviour near unity.
A striking feature of \ref{figure:pair_blendedness} is the qualitative difference seen in the distributions between N$_\text{max}=3$ and N$_\text{max}=4$.
Below this limit distributions tend to peak close to unity with long tails to higher values, as expected for a $\chi^2$ distribution, but above it the distribution is more evenly distributed, with no obvious peak near unity.
This strongly suggests that the (model-corrected) separations of pairs with N$_\text{max}\geq4$ have significant additional scatter between stars, well beyond what can be accounted for by statistical uncertainties.

Systematic effects on individual transitions due to weak blending are likely (at least partially) responsible for the star-to-star scatter seen.
We originally (and conservatively) chose to use only pairs with blendedness up to 2 based on a visual inspection of transitions in each category, and \ref{figure:pair_blendedness} provides evidence that this decision was well-motivated.
Indeed, from \ref{figure:pair_blendedness} even transitions with N$_\text{max}=3$ could potentially be used, but we chose to continue with our original selection limit to provide increased confidence in our results.
We note that the assigning of blendedness values is the one step in the analysis which we were not able to automate and which still requires human oversight, making the borders between categories necessarily somewhat imprecise.
It is difficult to tell visually whether a feature's shape is from weak blending or simply natural asymmetries due to photospheric convection, so it is possible that future work may be able to identify a more objective measure of blending.

\subsection{Absorption feature depth differences} \label{section:feature_depth_differences}

\begin{figure}
  \begin{center}
    \includegraphics[width=\linewidth]{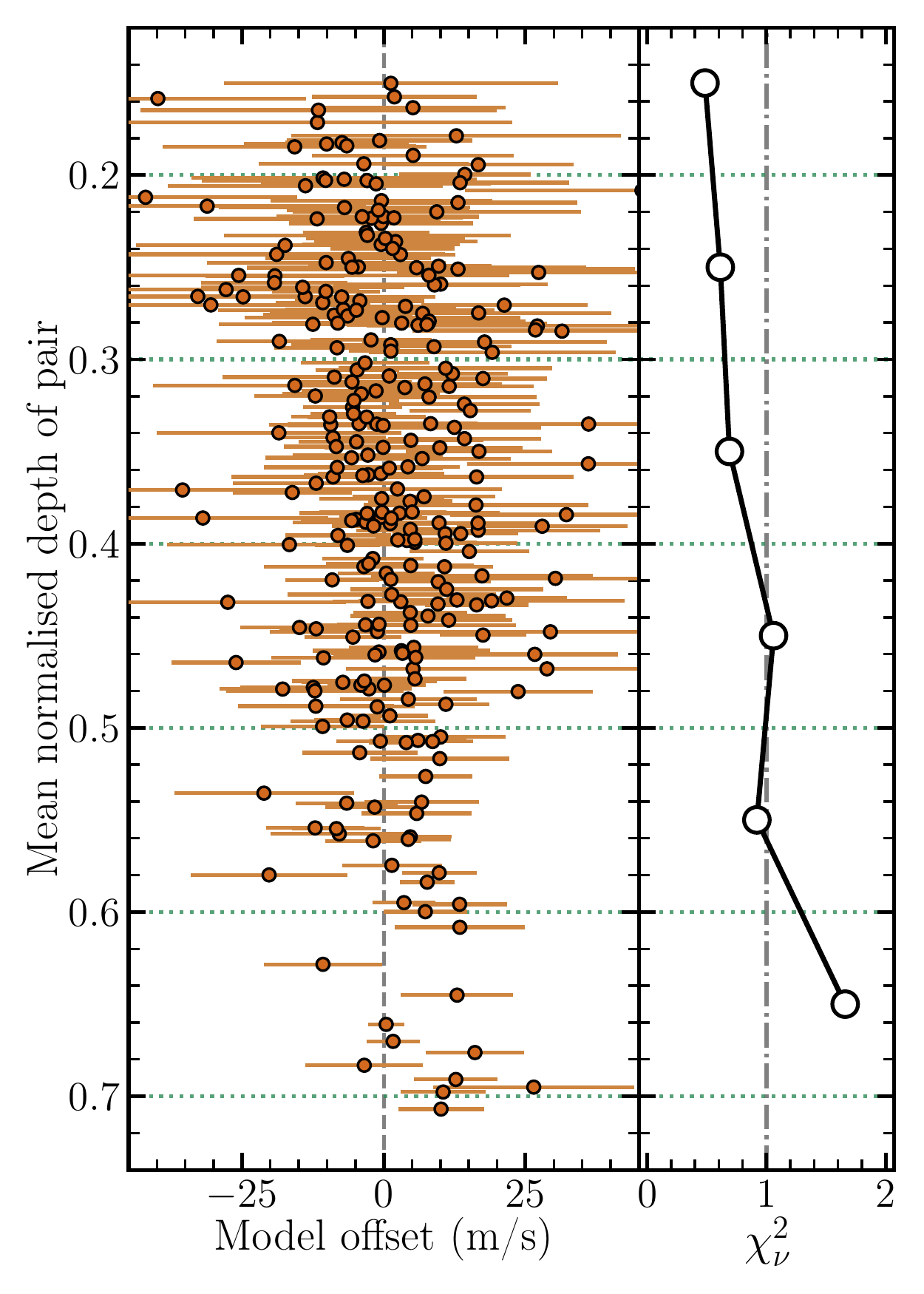}
    \caption{
    Example of how pair model offsets show slightly increased scatter for pairs with the deepest features.
    \textbf{Left panel:} Normalized mean depth of each pair in solar analogue HD~134060 plotted versus pair model offset.
    The uncertainties plotted are the quadrature sum of statistical uncertainties and \sigsys.
    \textbf{Right panel:} $\chi^2_\nu$ value of bins with a depth of 0.1 each.
    A slight increase in $\chi^2_\nu$ for the deepest pairs is seen in this star (in the lowest bin).
    }
    \label{figure:mean_depth_plot}
  \end{center}
\end{figure}

Differences in conditions in a star's photosphere with height impart similar asymmetries to features of similar normalized depths \citep{Dravins1982}.
We investigated two parameters to see if either correlated with increased scatter in pair model offsets in stars: the mean normalised depth of a pair's component features, and the difference in normalised depth between the two features.
\ref{figure:mean_depth_plot} shows an example of the weighted mean of pair model offsets plotted against mean normalised depth, for the star HD~134060.
The $\chi^2_\nu$ value is calculated in bins of 0.1 normalised depth to search for evidence of increased scatter as a function of depth.
HD~134060 shows a slight increase in scatter for the deepest pairs, which is representative of a few stars we checked with high [Fe/H] or low \Teff.
While the effects are small enough not to affect our results, it may be prudent to consider excluding transitions with absorption features deeper than $\sim$0.7--0.75 in future work.
However, we emphasis that this increase with depth is not seen in the majority of stars, which showed no discernible trends.
Similarly, no trends were discernible as a function of the difference in normalised depth between transitions in a pair, up to our chosen limit of a 0.2 difference.
While we did not investigate greater values for the difference, it may be useful for future work to investigate using a wider range of depths in pairs to increase the number of usable pairs.

\subsection{Charge transfer inefficiency} \label{section:charge_transfer_inefficiency}
Charge transfer inefficiency (CTI) refers to a systematic error found in CCDs due to the imperfect transfer of electrons between pixels during the CCD read-out process.
As each pixel on a CCD is read out sequentially at the end of an exposure, the accumulated charge is shifted from pixel to pixel with a small loss (typically $\ll1\%$) at each step.
This cumulative loss of charge can shift the measured wavelengths of emission features by causing changes in their shapes, and has been demonstrated in HARPS with a LFC by \citet{Zhao2021}.
Figure 5 in \citet{Zhao2021} shows the average shift (in units of pixel fraction) for emission features as a function of peak flux.
They caution that their results cannot be directly transferred from the emission lines of the LFC to absorption lines found in stars, but as the only estimate of CTI in HARPS we use them as a guideline to estimate CTI effects in our spectra.
CTI effects increase in relative magnitude with lower signal, due to the fractional nature of the charge loss at each step of the readout process.
In \citet{Zhao2021}, lower peak flux in an emission feature thus corresponds to a larger shift in that feature's wavelength.

If we assume that wavelength shift is solely a function of flux in a pixel (and not the surrounding distribution of flux as is likely the case in reality), we can treat the minimum flux in an absorption feature like the peak flux in an emission feature.
Due to the high SNR of our spectra, even the cores of most features have well above $10^4$ photons.
By comparison with figure 5 of \citet{Zhao2021}, we estimate a shift in the wavelength of features in our data of at most 0.01 pixel.
For HARPS at \SI{500}{\nano\meter} this corresponds to a shift of \SI{8.3}{\meter\per\second}.
At higher fluxes the shift measured in \citet{Zhao2021} decreases rapidly towards zero, so we may regard this value as a rough upper limit on the effects of CTI in our data.
We conclude that CTI effects are probably present in our data, but at a level below the typical statistical uncertainty (from our upper SNR limit) of \SI{\sim15}{\meter\per\second} for a single observation, and are thus negligible.

In some ways the effects of CTI are indistinguishable from those reported in \ref{section:cross_order_instances} related to pair separation variation across the CCD, though the effects we measured were large enough that they seem unlikely to be due to CTI alone.
We refer again to \ref{figure:duplicate_pairs} to emphasise that, whatever systematic effects may be present between the two sides of the CCD, our modeling of each pair is able to characterise these effects and effectively remove them.
We therefore leave a more detailed analysis and correction of CTI effects in HARPS to future work.

\subsection{Superimposed solar spectrum via scattered moonlight} \label{section:solar_lunar_reflection}
Sunlight reflected off the Moon and scattered in the Earth's atmosphere represents a possible source of systematic error in measuring transition wavelengths, and hence pair separations.
For Sun-like stars, a reflected solar spectrum is essentially a second, fainter copy of the target spectrum overlaid on it and offset from it by the radial velocity difference between the target and the Earth.
The relative intensity of this reflected solar spectrum -- and any consequent systematic offset in measured transition wavelengths -- is also dependent on the phase of the Moon and its proximity on the sky to the target.

\citet{Roy2020} used synthetic solar spectra with $R=100,000$ to measure the systematic errors in a spectral feature's wavelength from this effect.
They measured the wavelength shift as a function of two variables: the difference between target and background sky brightness (which encompasses proximity on-sky to the Moon), and the difference in radial velocity between the target and Earth.
They found that for the resolution used (close to HARPS's resolution $R\approx115000$), the shift is greatest at a certain critical radial velocity between the target and Earth, $|\Delta(\mathrm{RV})_\mathrm{crit}|\approx\SI{4}{\kilo\meter\per\second}$.
Their figure 4 shows the wavelength shift as a function of target and sky background brightness at $\Delta(\mathrm{RV})_\mathrm{crit}$.
We determine from it that the vast majority of stars in our sample are bright enough that, even if they were observed next to a full Moon, the systematic error from scattered moonlight is at most \SI{2}{\meter\per\second} and likely lower than \SI{0.5}{\meter\per\second} if observed in darker conditions.

One potential issue not considered in \citet{Roy2020} is thin cloud cover in the presence of scattered moonlight.
Such cloud cover will serve to both reduce the target signal and increase the sky background brightness.
Determining the degree of cloud cover for each observation in the sample proved infeasible, so we make the very conservative assumption of a full magnitude of target attenuation to determine what effect this would have.
By extrapolating the curves in figure 4 of \citet{Roy2020}, we estimate that the faintest stars in the sample (fainter than 7th magnitude) should still have less than \SI{20}{\meter\per\second} wavelength shift even if observed next to the full Moon.

We note again that to reach a shift of \SI{20}{\meter\per\second} for these faint stars would also involve the radial velocity difference at the time of observation being very close to $\Delta(\mathrm{RV})_\mathrm{crit}$, which is unlikely.
We plotted $\Delta(\mathrm{RV})$ versus the apparent magnitude of the target for every observation in the sample, and found that 44 observations from stars fainter than 7th magnitude fall within \SI{2.5}{\kilo\meter\per\second} of $\Delta(\mathrm{RV})_\mathrm{crit}$.
This represents a mere 0.43\% of the 10334 total observations in the sample.
As the fraction of observations potentially affected is so small, we accept that these observations may have an error of \SI{\sim20}{\meter\per\second}, without going to the step of calculating Moon phase and reconstructing cloud cover for each observation.
However, we stress that $>99$\% of observations in the sample should have an error from scattered moonlight of not more than \SI{\sim5}{\meter\per\second}, well below the noise floor for individual transitions.

\subsection{Cosmic rays and other non-Gaussian effects} \label{section:cosmic_rays}

\begin{figure}
  \begin{center}
    \includegraphics[width=\linewidth]{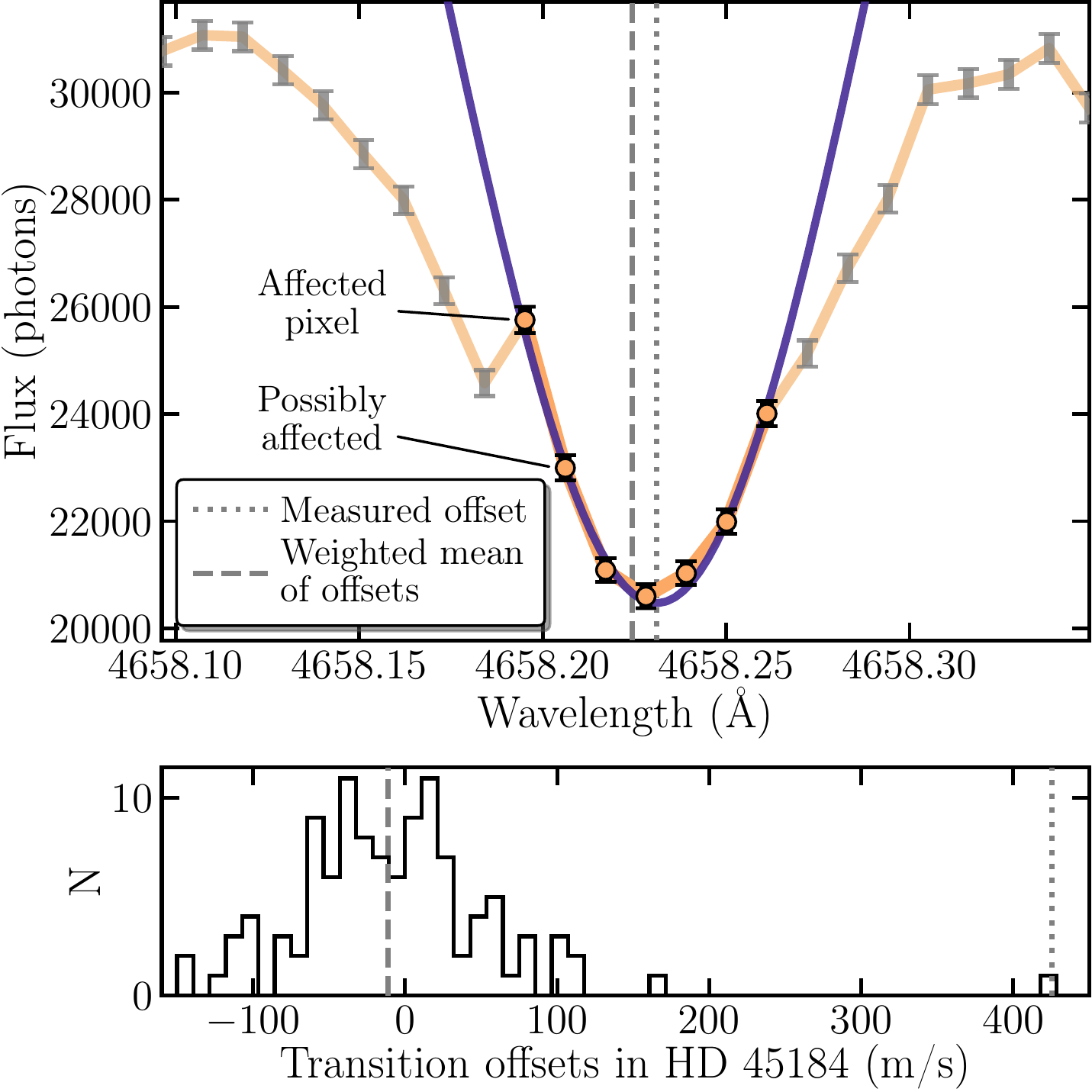}
    \caption{
    Example of cosmic ray event and effect on measured wavelength of feature.
    \textbf{Top panel:} An absorption feature from the transition \ion{Fe}{ii}~4658.285 in the star HD~45184 (gray error bars with solid line), from an observation taken 2012-05-10.
    A likely cosmic ray event is visible in the increased flux of the two left-most pixels of the seven central pixels used for fitting the feature (points with black error bars).
    The inverted Gaussian function (solid purple line) is the automated least-squares fit to the feature.
    \textbf{Bottom panel:} Distribution of the measured velocity offsets for this transition for all 111 observations of HD~45184.
    The offset in the top panel is $424\pm\SI{59}{\meter\per\second}$, a 7.4-$\sigma$ outlier from the weighted mean of \SI{-11}{\meter\per\second}.
    In both panels the dashed vertical line indicates the weighted mean and the dotted vertical line indicates the measured transition offset in the affected observation.
    The spectrum is not corrected for the radial velocity difference between Earth and HD~45184.}
    \label{figure:cosmic_ray_demonstration}
  \end{center}
\end{figure}

Cosmic rays impacting the HARPS detector during observations have the potential to introduce deviations in the measured wavelengths of absorption features.
If one or more of the pixels impacted by the cosmic ray is in the core of an absorption feature of interest, the shape of the feature (and thus our measured wavelength) can be affected.
No cosmic ray cleaning is involved in the creation of the e2ds files used in this work.
\ref{figure:cosmic_ray_demonstration} shows an example of a likely cosmic ray hit in a single observation of the transition \ion{Fe}{ii}~4658.285 in the solar twin HD~45184 taken on 2012-05-10.
We note that there is no trace present of such a single-pixel `spike' in any of the other observations of this feature in HD~45184, and that this is clearly an unusual event.
The automated fitting procedure uses the three pixels on either side of the deepest pixel detected (as discussed in \ref{section:measuring_pair_separations}), and the cosmic ray has significantly increased the measured flux in (at least) the left-most two pixels.
This has resulted in a transition velocity offset measurement of \SI{426}{\meter\per\second} from the expected position with formal statistical error of \SI{59}{\meter\per\second}.
The weighted mean and RMS of the distribution of velocity offsets for this transition in HD~45184, based on 111 observations, are \SI{-11}{\meter\per\second} and \SI{71}{\meter\per\second}, making this wavelength measurement a 7.4-$\sigma$ outlier.
This measurement is accordingly excluded by the simultaneous outlier rejection and stellar parameter fitting process described in \ref{section:transition_wavelength_variation}.

The overall effect of cosmic rays is to introduce an additional non-Gaussian spread in the distribution of wavelengths measured for individual transitions.
Rather than attempt any sort of cosmic-ray cleaning procedure on the data ourselves, we instead added the sigma-clipping functionality detailed in \ref{section:transition_wavelength_variation} to the process of fitting transition measurements as a function of stellar parameters.
Outliers more than $3\sigma$ are excluded from use in forming pair separation measurements, which should remove even mild instances of this effect.
While cosmic rays are the only source of non-Gaussian effects we explicitly consider here, any other causes of such effects should be similarly dealt with by our analysis process.

\subsection{Stellar activity cycles} \label{section:magnetic_activity_cycles}
The Sun has a well-studied magnetic activity cycle with a period of approximately 11 years, and it is likely that other Sun-like stars will have similar cycles.
Indeed, several stars in the sample have cycles with periods estimated from chromospheric activity \citep{BoroSaikia2018}.
Magnetic activity could potentially alter the shapes of absorption features and change measured pair separations.
This effect would manifest as a greater scatter, compared to the statistical uncertainties, in the pair separation measurements of a single star, and may also cause a systematic shift compared to other stars, especially if discovered to be variable over the course of one or more periods of activity.

To search for evidence of extra scatter caused by magnetic activity, we focused upon two solar twins, with published activity periods, for which we have the largest number of observations: HD~45184 and HD~146233, with 111 and 139 observations, respectively.
The temporal baselines over which these observations were taken are 13.3 and 12.6 years and \citet{BoroSaikia2018}, using a collection of historical measurements from Mount Wilson Observatory and HARPS, list tentative activity periods for these stars of $4.9\pm0.3$ and $11.4\pm1.2$ years, respectively.
\citet{Baum2022} used additional historical observations to confirm that HD~146233 has a variable cycle period, which is seen to be well-sampled from $\sim$1992--2020 in their figure 3.
While the cycle of HD~146233 may be variable, this figure confirms that there was a change of $\sim$12\% in the measure of its magnetic activity over the temporal baseline of the HARPS observations we use here.
Lacking similar information about HD~45184, we assume that its quoted cycle of $4.9\pm0.3$ years is correct and note that our temporal baseline for it is over twice this length, making it probable that the star underwent at least one magnetic cycle during that time even if its period is variable.

For the 17 pairs of transitions used in \citetalias{Berke2022a}, we inspected distributions of the pair model offsets and their constituent transition model offsets in both stars as functions of time.
No visual evidence of scatter significantly above the statistical level of $\sim$\SI{15}{\meter\per\second} implied by the statistical uncertainties for the observations was found for any of the pairs.
The $\chi^2_\nu$ values of the distributions similarly showed no significant excess scatter for any of the 17 pairs.
We have focused on these two solar twins with the most observations and estimated cycle periods as offering the best opportunity to observe scatter from magnetic activity and, similarly, we have focused on the pairs we can currently use to constrain \varalpha.
Future work with additional stars or pairs may discover an additional error component which can be modelled and removed, but we conclude that it likely does not contribute a significant amount of systematic error at our level of precision.

\subsection{Transiting exoplanets}
The transit of a smaller star in front of a larger primary changes the apparent radial velocity of the primary due to the Rossiter--McLaughlin effect \citep{Rossiter1924, McLaughlin1924}, which has also been observed for transiting exoplanets \citep[first reported in][]{Queloz2000}.
The shapes of spectral features in the primary star are modified due to the companion alternately blocking blue- or red-shifted light as it transits different hemispheres of the rotating primary.
Observing a star while a planetary transit is in progress could thus potentially cause deviations in the measured wavelengths of features.
Achromatic Doppler shifts of the entire spectrum, from either transiting or non-transiting planets, should not cause any effect due to our use of differential pair separations.
However, pair separations could potentially be affected by differential changes in the shapes of absorption features due to the varying range of scale heights in the photosphere over which they are formed.

A total of 31 stars in the stellar sample are known to host exoplanets.
We searched for transiting planets among these stars by checking each one in the Extrasolar Planets Encyclopaedia\footnote{\url{http://exoplanet.eu/}} for the presence of a planetary radius measured via transit lightcurve.
Two stars, HD~39091 (N$_\text{obs}=48$) and HD~136352 (N$_\text{obs}=266$), were found to have transiting planets.
In a similar manner to searching for evidence of an effect from stellar magnetic activity cycles (\ref{section:magnetic_activity_cycles}), we visually inspected the model offsets of pairs (and their component transitions) for the 17 pairs used in \citetalias{Berke2022a} for signs of excess scatter in both stars.
We found no indication of additional scatter or outliers that might be due to transiting planets, and conclude that there is unlikely to be any effect above the statistical \SI{\sim15}{\meter\per\second} scatter level expected from the uncertainties on individual features.
While the shift in individual features due to the Rossiter--McLaughlin effect is extremely difficult to predict \emph{a priori} without detailed knowledge of the system, we note for comparison that in \citet{Queloz2000} a change of approximately \SI{\pm35}{\meter\per\second} was measured in the radial velocity of the primary using cross-correlation.
However, the differential effect on the separations of pair of transitions is likely to be considerably smaller than this bulk shift.

\subsection{Differing elemental abundances and isotope ratios} \label{section:elemental_abundances_and_isotope_ratios}
\begin{figure}
  \begin{center}
    \includegraphics[width=\linewidth]{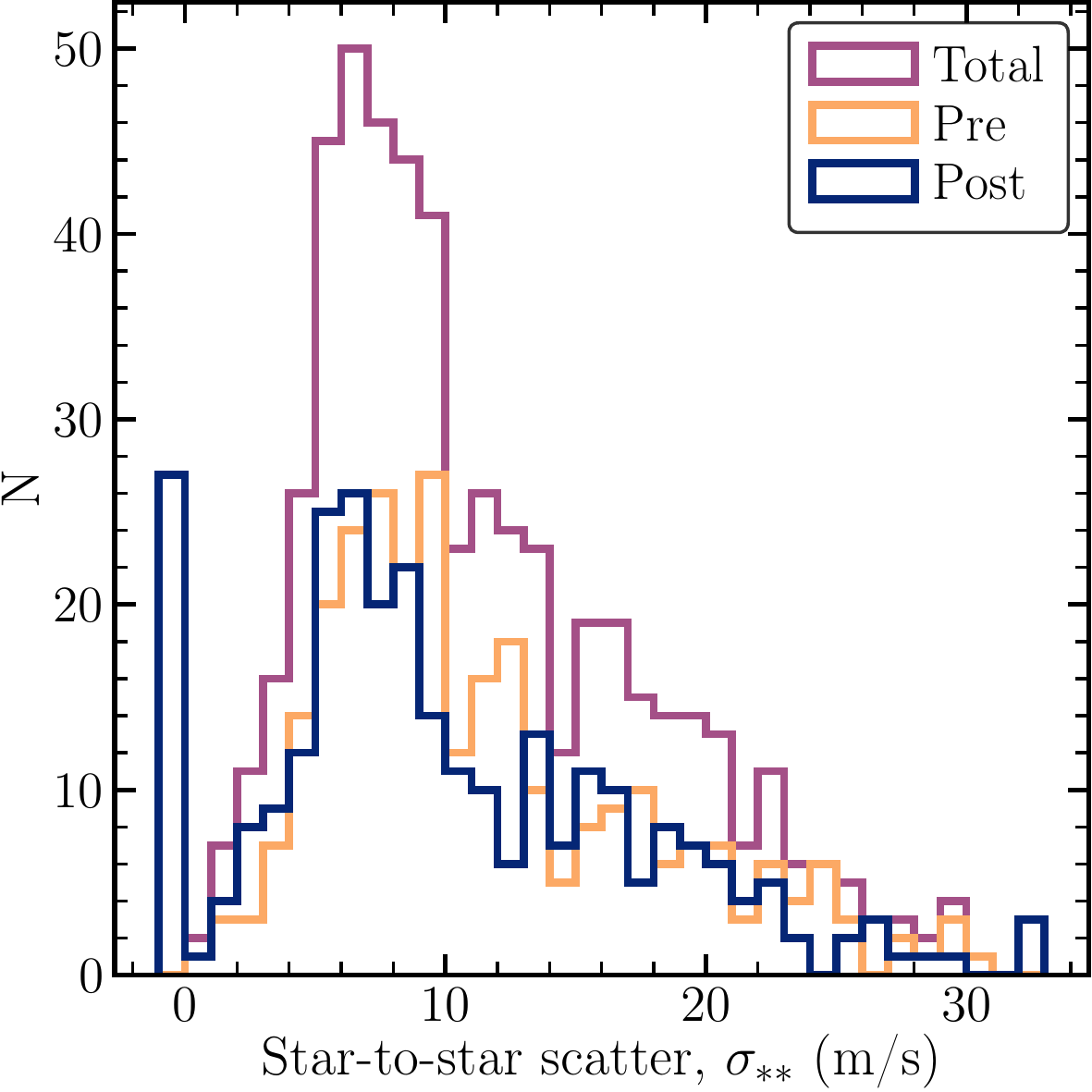}
    \caption{The distribution of \sigsys values for the 284 pairs in the sample.
    We distinguish between the `pre' and `post' observations, with the `total' being the sum of the two histograms.
    The `pre' and `post' distributions have medians of \SI{9.9}{\meter\per\second} and \SI{8.4}{\meter\per\second}, respectively.
    The left-most bin indicates a value of zero, i.e. pairs for which no additional scatter term was required to lower $\chi^2_\nu$ for the pair to unity; only observations in the `post' group fall into this bin.
    }
    \label{figure:sigma_s2s_histogram}
  \end{center}
\end{figure}

As described in \ref{section:pair_separation_variation}, we discovered a systematic noise floor between stars for all transition pairs in pre-fibre change observations, and 90\% of pairs in post-change observations.
We use the term \sigsys to denote this level of star-to-star scatter and consider its possible origins here.
In \ref{figure:sigma_s2s_histogram} histograms of the \sigsys values for pairs in the sample are shown, for observations taken before and after the HARPS optical fibre change, and the sum of both histograms.
Note the existence of a gap between zero and the peak in both the pre and post distributions' (non-zero) values.
This indicates that the non-zero \sigsys values are real, and not a statistical artefact.
If a pair's separation values were normally distributed, the $\chi^2_\nu$ value for the pair would randomly fall above or below one with roughly equal frequency.
In turn, \sigsys would be zero or non-zero equally often, and the plot would resemble half a Gaussian function with half the values in the left-most bin at zero and the rest tailing off normally.
The gap between the peak and zero in both distributions clearly shows this is not the case.

We performed the two-sample Kolmogorov-Smirnov (KS) test on the pre and post samples to check if they could have been drawn from the same underlying distribution.
Both samples have 284 data points, treating instances of the same pairs separately.
The formal KS test statistic for the two samples is 0.158, with a $p$-value of 0.0016.
Given the $p$-value of less than 1\%, we can be reasonably confident in rejecting the null hypothesis that both samples are drawn from the same underlying distribution, validating our decision to treat observations from the two observing eras independently.

The most obvious difference in the post sample compared to the pre sample is the presence of 27 pairs (9.5\%) with $\sigsys=0$, indicating that no additional scatter term is necessary.
The shape of the post distribution supports the conclusion that the extra scatter for most pairs is a combination of astrophysical and instrumental factors.
Prior to the HARPS fiber change, every pair in the sample required the additional scatter term \sigsys, while after the change, 9.5\% of pairs in the sample no longer required it.
As reported in \citet{LoCurto2015}, the overall calibration accuracy of HARPS improved after the fibre change.
The post distribution thus suggests that, for at least some pairs, any extra star-to-star scatter seen prior to the fibre change was a result of small-scale instrumental calibration distortions.
However, over 90\% of pairs, even after the fiber change, retain a non-zero \sigsys.
We cannot rule out that this may be due to remaining distortions in HARPS's calibration, but the fact that the two pre and post distributions have very similar shapes, well-separated from zero, suggests that there is a real astrophysical component to \sigsys as well.

One possible explanation for this star-to-star scatter lies in differing elemental abundances between stars.
We use the abundance of iron, [Fe/H], as a proxy for overall metallicity in accordance with common usage, but did not investigate the elemental abundances of stars in the sample in greater detail.
Even a small change in elements other than iron could potentially produce the small star-to-star differences seen here via otherwise-undetectable weak blends.
\ref{figure:transition_density} illustrates the typical density of known transitions and the potential for weak blends to shift absorption feature shapes enough to have a measurable effect.
From \ref{equation:feature_velocity_shift} we can estimate a typical feature velocity shift.
For a hypothetical feature of interest and weak secondary feature with normalized depths $D_1, D_2=0.5, 0.001$ (50\% and 0.1\% of the continuum) and a separation of \SI{3}{\kilo\meter\per\second}, \ref{equation:feature_velocity_shift} predicts a shift in the main feature of \SI{6}{\meter\per\second}.
This falls within the range of \sigsys values seen in both pre and post distributions, and supports the possibility that weak blends from different elemental abundances could be responsible for the observed star-to-star scatter.
Nevertheless, we would expect this effect to be suppressed, to some degree, by the comparison of pair separations between stars.

Another possible explanation lies in differing isotope ratios between stars.
Isotope shifts are typically small, on the order of a few hundred \si{\meter\per\second}, and thus unresolved.
For elements with multiple stable isotopes, the wavelength measured for a transition is really a combination of contributions from each isotope.
Differences in isotope ratio between stars could therefore cause changes in pair separations and contribute to \sigsys.

Unfortunately, there is a paucity of information in the literature regarding isotope shifts for transitions in the visible regime, or on isotope ratio differences between stars in the Milky Way.
However, we can make an order of magnitude estimate based on available information as follows.
The Kurucz list (see \ref{section:transition_selection}) contains information on the isotope shift for one of the transitions in our sample, the \ion{Ni}{i} transition at \SI{4521.249}{\angstrom}.
The wavelengths of this transition in the two most abundant isotopes of nickel (\BPChem{\^{58}Ni} and \BPChem{\^{60}Ni}) have a difference of \SI{6}{\milli\angstrom}, or just under \SI{400}{\meter\per\second}.
If we assume a toy model of a transition from an element with just these two isotopes in the abundance ratio 3:1, then its wavelength will be shifted by \SI{\sim100}{\meter\per\second} from that of the dominant isotope.
If we then assume a second star with an isotope ratio of 2:1, the centroid shift would instead be \SI{\sim130}{\meter\per\second}, with the star-to-star scatter between the two therefor being \SI{\sim30}{\meter\per\second}.
Such isotope ratio differences could thus plausibly account for the star-to-star scatter we observe.

Little information on stellar isotope ratio variation in the Milky Way is available in the literature,
but \citet{Yong2003} show that the magnesium isotopic ratio, \BPChem{\^{24}Mg:\^{25}Mg:\^{26}Mg}, varies between 100:0:0 and 60:20:20 in a sample of 61 cool dwarfs and giants.
Their stellar sample was generally cooler and more metal-poor than ours and we use it only for comparison (being one of the few isotopic ratio measurements available), but the isotope ratio differences seen are comparable in magnitude to the toy model used above.
While much work remains to be done on isotope ratio variation between stars, we find it plausible that they could account for the excess star-to-star scatter in pair separations we observe (possibly in tandem with elemental abundance changes).

\subsection{Stellar rotational velocity} \label{section:stellar_rotational_velocity}
An additional possible cause of the residual differences between stars is the projected rotational velocity, $V\sin{i}$.
\citet{Smith1987} synthesised absorption profiles for several transitions in the Sun and, contrary to expectations, discovered that increased rotation speed actually enhanced the asymmetries already present due to convection.
While the cores of features were less affected than the wings (part of our motivation for only fitting the cores), even the cores were shifted (by different amounts) for each transition, potentially up to tens of \si{\meter\per\second}.
The strength of the effect was also noted to be a non-linear function of the rotational velocity, being roughly quadratic for a rotational velocity below $\SI{20}{\kilo\meter\per\second}$.
\citet{Dravins1990} confirmed this behaviour with several additional synthesised absorption features computed for a range of temperatures around the solar temperature.
In both works, increasing rotational velocity led to increased asymmetry (and corresponding centroid shifts), but with an amplitude that differed depending on aspects of the transition such as its depth.

Determining $V\sin{i}$ for all stars in the stellar sample was outside the scope of this work, so we simply note here that differences in it may also (along with the sources discussed in \ref{section:elemental_abundances_and_isotope_ratios}) account in part for the residual star-to-star scatter observed.
Through visual inspection of spectra we can say that for all stars in the stellar sample the rotational velocity is not obviously much greater than for the Sun (\SI{2}{\kilo\meter\per\second})\footnote{Two stars, excluded for other reasons, had noticeably broader spectral features, so the difference was clearly visible.}.
Still, the non-linearity of the effect and the observed difference in features of different depths \citep{Smith1987,Dravins1990} makes the determination of the exact effect on any given transition across the sample difficult to predict.
However, this dependence on $V\sin{i}$ may make it possible to disentangle its effects on transitions for an individual star and remove them in the future.

\subsection{Other possible effects} \label{section:other_possible_effects}
While we have attempted to consider the most likely causes of systematic effects in the solar twins approach above, it certainly remains possible that other stellar astrophysical effects should be considered in future. However, the consistency of the velocity separation measurements in \citetalias{Berke2022a} between stars emphasises that the cumulative effect of all systematic errors must be subtle and not obvious or easily detected even with the larger sample of stars studied in this paper (which cover a wider range of stellar parameters than those in \citetalias{Berke2022a}). At present, the only systematic effects, which appear random from star to star, are adequately quantified by the \sigsys term for each pair. Future work may be able to better determine the cause of these intrinsic differences between stars, but for purposes of this paper it is enough to quantify their effects on the systematic error budget via \sigsys

\section{Conclusions} \label{section:conclusions}
The solar twins method, first introduced in \citet{Murphy2022b} and detailed in \citet{Berke2022a} (\citetalias{Berke2022a}, a companion paper to this paper), constrains variation in the fine-structure constant by comparing separations between pairs of transitions across very similar stars, rather than to an absolute laboratory reference.
This paper provides the first investigation into a wide array of possible systematic errors in the solar twins method: wavelength calibration distortions in HARPS, systematic changes in stellar line wavelengths as a function of their atmospheric parameters, blending of stellar lines, normalized line depths, separation between transitions in a pair, charge transfer inefficiency, scattered sunlight reflected off the Moon, cosmic ray impacts, stellar magnetic activity cycles, transiting exoplanets, differing elemental abundances and isotope ratios, and stellar rotational velocity.
The solar twins method ultimately proves robust when extended to the range of solar analogues (as defined in \ref{equation:star_definitions}), and we find just a single dominant systematic error in the form of small ($\lesssim\SI{30}{\meter\per\second})$ intrinsic star-to-star differences.
We discuss these main results further below.

For this investigation we used a sample of 130 bright, nearby stars selected for their similarity to the Sun in terms of their stellar atmospheric parameters \Teff, [Fe/H], and $\log{g}$.
The final sample covered the parameter ranges $\SI{5413}{\kelvin}\leq\Teff\leq\SI{6077}{\kelvin}$, $-0.44\leq\text{[Fe/H]}\leq0.40$, and $4.06\leq\log{g}\leq4.56$.
A total of 10126 observations with $\text{SNR}\geq200$ from the HARPS archive, from the years 2004--2017, contributed to this work.
Transitions were carefully selected to avoid telluric lines (down to the level of 0.1\% of the continuum), and we classified each one in terms of blending with stellar features.
We checked for systematic errors as a function of `blendedness', and found that, while scatter in pair separations increased with increasing blendedness, our selection for this paper was conservative and could even potentially be expanded safely in future.
The separations between 284 pairs of transitions were modelled as a multi-variate quadratic function of the three stellar parameters mentioned above, providing a reference -- defined by the stars themselves -- against which each star can be compared.
This technique allows differences in pair separation down to the level of a few \si{\meter\per\second} to be measured.

The main finding of this paper is that the range of applicability of the solar twins method can be extended beyond solar twins to solar analogues, stars with $\Teff\pm\SI{300}{\kelvin}$, $\text{[Fe/H]}\pm0.3$, and $\log{g}\pm0.4$ around solar values (though we continue to use the name `solar twins method' for simplicity).
This represents a doubling or trebling of the allowed range for each of the parameters, with an attendant increase in the number of stars which can be used.
This will be important for future studies of \varalpha on the Galactic scale, where constraints in source selection precision make identifying solar twins more difficult.
Initial progress in this direction is described in \citet{Lehmann2022}, and future results will be reported in \textcolor{black}{Lehmann et al. (in prep.) and Liu et al. (in prep.)}.
The models of pair separation (as a function of stellar parameters) are also important outcomes of this paper, since they serve as the reference values against which stars in the range of applicability can be compared.
A similar modelling procedure applied separately to transitions also allows us to reject spurious measurements from individual stellar exposures.
This leads to very robust results, with no sign of needing to remove additional measurements.
As shown in \ref{section:pair_separation_variation}, pair separations vary in a simple way with the atmospheric parameters \Teff, [Fe/H], and $\log{g}$, which would impart systematics of up to hundreds of \si{\meter\per\second} if not corrected.

Another important result of this paper is the presence of a non-zero star-to-star scatter, denoted \sigsys.
This value -- the dominant systematic effect currently identified for $>90\%$ of pairs -- is an extra scatter term which brings the $\chi^2_\nu$ value for a pair down to unity when added in quadrature to the statistical error, i.e. it is the excess scatter beyond that expected from the uncertainties.
As shown in \ref{figure:sigma_s2s_histogram}, the distribution of \sigsys values for individual pairs ranges from 0 to \SI{33}{\meter\per\second}, with a median of \SI{9}{\meter\per\second}.
Having observations from before and after the HARPS fibre change in 2015 provides useful information: prior to the change 100\% of pairs had a non-zero \sigsys, while afterwards 9.5\% of pairs no longer needed the extra scatter term.
This indicates that some fraction of \sigsys is likely to be instrumental, since the overall stability of HARPS's wavelength calibration reportedly improved slightly after the change \citep{LoCurto2015}.
However, the very similar shapes of the distributions for the pre- and post-change observations also indicates that there is most likely an astrophysical component as well.
The fact that both pre and post distributions peak at positive values with a sharp drop-off towards zero implies that these star-to-star scatters are indicative of real differences between stars.
Although the exact origin of these differences is as-yet unknown, we estimate in \ref{section:elemental_abundances_and_isotope_ratios} that differences in elemental abundance or isotope abundance ratios between stars provide a plausible explanation.
Simple, toy-model considerations indicate that weak, otherwise-undetectable lines from elements or isotopes blending with absorption lines of interest could create the star-to-star differences observed.
It is important to note that any other effects that contribute to the star-to-star scatter but not explicitly considered here are still included in our error budget.

Future work may be able to further extend the range of parameters over which the solar twins method can be applied.
As discussed in \ref{section:stellar_parameter_dependence}, we chose to limit the range of temperatures and metallicities we ultimately used based on the discovery of increasing star-to-star scatter in many pairs beyond the solar analogue limits.
This increased scatter could instead potentially be modelled, allowing an even wider of range of stars to be compared.
Alternatively, the same techniques of the solar twins method could be applied to other types of stars, not including the range of stellar parameters around the Sun.
Stars intrinsically more luminous than solar twins -- for example, giants \citep{Hees2020} -- would have the obvious advantage of providing high-SNR spectra at greater distances.

This paper has demonstrated that solar analogues can serve as viable probes for constraining variation in $\alpha$, with appropriate care and accounting for systematic errors.
Indeed, the use of solar analogues allows for levels of precision nearly two orders of magnitude more precise than other probes previously used in astronomical searches \citep[e.g., quasar absorption systems,][]{Murphy2017}.
This opens up the Galactic scale to varying-$\alpha$ searches with unprecedented precision, in turn allowing for testing for any variation in $\alpha$ against parameters such as dark matter density.

\subsection*{Acknowledgements}
We thank Dainis Dravins for discussions about potential astrophysical systematic errors.

\textit{Funding:} DAB, MTM and FL acknowledge the support of the Australian Research Council through \textsl{Future Fellowship} grant FT180100194.

\textit{Facilities.} This research has made use of data or services obtained from, or tools provided by: the Extrasolar Planets Encyclopaedia (Jean Schneider, CNRS/LUTH - Paris Observatory), the SIMBAD database, operated at CDS, Strasbourg, France, NASA's Astrophysics Data System, the Geneva--Copenhagen survey \citep{Nordstrom2004}, the Belgian Repository of fundamental Atomic data and Stellar Spectra \citep{Lobel2008}, atomic line lists compiled by R. Kurucz, the NIST Atomic Spectra Database \citep{Kramida1999, Kramida2013}, the Transmissions Atmosph\'eriques Personnalis\'ees Pour l'AStronomie project \citep{Bertaux2014}, and the ESO Science Archive Facility.

\textit{Software.} This research has made use of: Python \citep{VanRossum1995}, NumPy \citep{Harris2020}, SciPy \citep{Virtanen2020}, Astropy \citep{Astropy2013, Astropy2018}, Astroquery \citep{Ginsburg2019}, CMasher \citep{vanderVelden2020}, IPython \citep{Perez2007}, \texttt{matplotlib} \citep{Hunter2007}, \texttt{unyt} \citep{Goldbaum2018}, \texttt{hickle} \citep{Price2018}, \texttt{tqdm} \citep{DaCostaLuis2021}, TOPCAT \citep{Taylor2005}, \texttt{ds9} \citep{Joye2003}, and \texttt{pandas} \citep{McKinney2010, McKinney2011}.

%%%%%%%%%%%%%%%%%%%%%%%%%%%%%%%%%%%%%%%%%%%%%%%%%%
\subsection*{Data Availability}
Based on observations obtained from the ESO Science Archive Facility and collected at the European Southern Observatory under ESO programmes 072.C-0488(E), 183.C-0972(A), 192.C-0852(A), 196.C-1006(A), 198.C-0836(A), 077.C-0364(E), 60.A-9036(A), 091.C-0936(A), 188.C-0265(C), 188.C-0265(E), 188.C-0265(G), 188.C-0265(J), 188.C-0265(K), 188.C-0265(O), 188.C-0265(Q), 075.C-0332(A), 183.D-0729(A), 079.D-0075(A), 074.C-0012(B), 083.C-1001(A), 084.C-0229(A), 085.C-0318(A), 086.C-0230(A), 089.C-0050(A), 090.C-0849(A), 092.C-0579(A), 093.C-0062(A), 094.C-0797(A), 095.C-0040(A), 096.C-0053(A), 097.C-0021(A), 077.C-0530(A), 079.C-0681(A), 188.C-0265(A), 188.C-0265(P), 188.C-0265(R), 074.D-0380(A), 075.D-0614(A), 082.C-0315(A), 087.C-0990(A), 088.C-0011(A), 188.C-0265(M), 076.C-0878(B), 078.C-0833(A), 091.C-0271(A), 188.C-0265(B), 188.C-0779(A), and 088.C-0323(A).
The raw data for this work is available from the ESO Science Archive facility, at \url{http://archive.eso.org/cms/eso-data/eso-data-direct-retrieval.html}.


%\bibliographystyle{mnras}
%\bibliography{Astrophysics.bib}

%%%%%%%%%%%%%%%%%%%%%%%%%%%%%%%%%%%%%%%%%%%%%%%%%%

%%%%%%%%%%%%%%%%% APPENDICES %%%%%%%%%%%%%%%%%%%%%

%\appendix

%%%%%%%%%%%%%%%%%%%%%%%%%%%%%%%%%%%%%%%%%%%%%%%%%%

% Don't change these lines
\bsp	% typesetting comment
\label{lastpage}
\end{CJK*}
\end{document}